\documentclass[a4paper]{article}%
\usepackage{amssymb}
\usepackage{amsmath}
\usepackage{amstext}
\usepackage{amsthm}
\usepackage{xspace}
\usepackage[dvips]{graphicx}
\usepackage{amsfonts}%
\providecommand{\U}[1]{\protect\rule{.1in}{.1in}}

\newtheorem{theorem}{Theorem}
\newtheorem{lemma}[theorem]{Lemma}
\newtheorem{prop}[theorem]{Proposition}
\newtheorem{proposition}[theorem]{Proposition}
\newtheorem{cor}[theorem]{Corollary}

\theoremstyle{definition}
\newtheorem{definition}[theorem]{Definition}
\newtheorem{rem}[theorem]{Remark}
\theoremstyle{remark}

\numberwithin{equation}{section}
\begin{document}

\title{Hylomorphic solitons}
\author{Vieri Benci\\Dipartimento di Matematica Applicata \textquotedblleft U.
Dini\textquotedblright\\Universit\`{a} di Pisa\\Via Filippo Buonarroti 1/c, 56127 Pisa, Italy\\e-mail: benci@dma.unipi.it}
\maketitle

\begin{abstract}
This paper is devoted to the study of solitary waves and solitons whose
existence is related to the ratio energy/charge. These solitary waves are
called hylomorphic. This class includes the Q-balls, which are spherically
symmetric solutions of the nonlinear Klein-Gordon equation (NKG), as well as
solitary waves and vortices which occur, by the same mechanism, in the
nonlinear Schroedinger equation and in gauge theories. This paper is devoted
to the study of hylomorphic soliton. Mainly we will be interested in the very
general principles which are at the base of their existence such as the
Variational Principle, the Invariance Principle, the Noether theorem, the
Hamilton-Jacobi theory etc. We give a general definition of hylomorphic
solitons and an interpretation of their nature (\textit{swarm interpretation})
which is very helpful in understanding their behavior. We apply these ideas to
the Nonlinear Schroedinger Equation (NS) and to the Nonlinear Klein-Gordon
Equation (NKG) repectively.

\end{abstract}
\tableofcontents

\section{Introduction}

Roughly speaking a solitary wave is a solution of a field equation whose
energy travels as a localized packet and which preserves this localization in
time. A \textit{soliton} is a solitary wave which exhibits some strong form of
stability so that it has a particle-like behavior.

\ 

Today, we know (at least) three mechanism which might produce solitary waves
and solitons:

\ 

\begin{itemize}
\item Complete integrability; e.g. Kortewg-de Vries equation
\[
u_{t}+u_{xxx}+6uu_{x}=0
\]

\item Topological constrains: e.g. Sine-Gordon equation
\[
u_{tt}-u_{xx}+\sin u=0
\]

\item Ratio energy/charge: e.g. the following nonlinear Klein-Gordon equation
\[
\psi_{tt}-\Delta\psi+\frac{\psi}{1+\left\vert \psi\right\vert }=0;\;\psi
\in\mathbb{C}%
\]

\end{itemize}

This paper is devoted to the third type of solitons which will be called
hylomorphic solitons. This class of solitons includes the $q$-balls (see
\cite{Coleman86}) which are spherically symmetric solutions of
as well as solitary waves and vortices which occur, by the same mechanism, in
the nonlinear Schroedinger equation and in gauge theories.

The most general equations for which it is possible have hylomorphic solitons
need to have the following features:\label{assA}

\begin{itemize}
\item \textbf{A-1. }\textit{The equations are variational namely they are the
Euler-Lagrange equation relative to a Lagrangian density $\mathcal{L}$}.

\item \textbf{A-2. }\textit{The equations are invariant for time translations,
namely $\mathcal{L}$ does not depend explicitly on }$t.$

\item \textbf{A-3. }\textit{The equations are invariant for a gauge action,
namely $\mathcal{L}$ does not depend explicitly on the phase of the field
}$\Psi$ \textit{which is supposed to be complex valued (or at lest to have
some complex valued component).}
\end{itemize}

By Noether theorem \textbf{A-1} and \textbf{A-2} guarantee the conservation of
energy, while \textbf{A-1} and \textbf{A-3 }guarantee the conservation of an
other integral of motion which we call \textit{hylenic charge}.

The existence of hylomorphic solitons is guaranteed by the interplay between
energy and hylenic charge\textit{.}

This paper is devoted to the study of hylomorphic soliton. Mainly we will be
interested in the very general principles which are at the base of their
existence such as the Variational Principle, the Invariance Principle, the
Noether theorem, the Hamilton-Jacobi theory etc. These principles will be
discussed in section \ref{GT}. We will also give a proof of Noether theorem in
the form needed for this study.

In section \ref{HS}, we give a general definition of hylomorphic solitons and
we give an interpretation of their nature (\textit{swarm interpretation})
which is very helpful in understanding their behavior.

In sections \ref{SSE} and \ref{SKG}, we apply these ideas to the Nonlinear
Schroedinger Equation (NS) and to the Nonlinear Klein-Gordon Equation (NKG)
repectively. The amount of results relative to NSE end NKG is huge (we just to
cite some of the main papers: \cite{D64}, \cite{rosen68}, \cite{strauss},
\cite{coleman78}, \cite{BL81}, \cite{CL82}, \cite{shatah}, \cite{Coleman86}
and the books \cite{rub}, \cite{yangL})); however, here we restrict ourselves
to the aspects of these equations related to the ideas of the previous
sections; in particular, we will emphasize their common "hylomorphic" nature
which has been recently discovered. The technical aspects related to the study
of these equations will be treated superficially, but we will refer the
interested readers to the appropriate papers.

An other set of equations to which this kind of ideas can been applied are the
Klein-Gordon-Maxwell Equations (NKGM) and also more general gauge theories
such as the Yang-Mills equations. In this paper, we will not discuss these
equation and we refer to \cite{BF02}, \cite{sammomme}, \cite{ca}, \cite{tea},
\cite{tea2}, \cite{ingl}, \cite{befogranas}, \cite{BF09TA} and their references.

An other very interesting aspect of NS, NKG and NKGM is the fact that they
admit \textit{hylomorphic vortices }namely solitary waves having a
nonvanishing angular momentum. The existence of these vortices is guarateed by
a mechanism similar to that of hylomorphic solitary waves. Here we will not
discuss the \textit{hylomorphic vortices }and we refer to \cite{bevi},
\cite{befov07},\textit{\ }\cite{befogranas}, \cite{bebo} and \cite{BF09TA}.

\section{The general theory\label{GT}}

\subsection{The variational principle\label{212}}

The fundamental equation of Physics are the Euler-Lagrange equations of a
suitable functional. This fact is quite surprising. There is no logical reason
for this. It is just an empirical fact: all the fundamental equations which
have been discovered until now derive from a variational principle.

For example, the equations of motion of $k$ particles whose positions at time
$t$ are given by $x_{j}(t),\;x_{j}\in\mathbf{R}^{3},\;$ $j=1,...,k$ are
obtained as the Euler Lagrange equations relative to the following functional
\begin{equation}
\mathcal{S}=\int\left(  \sum_{j}\frac{m_{j}}{2}\left|  \dot{x}_{j}\right|
^{2}-V\left(  t,x_{1},.....,x_{k}\right)  \right)  dt \label{L1}%
\end{equation}
where $m_{j}$ is the mass of the $j$-th particle and $V$ is the potential
energy of the system.

More generally, the equations of motion of a finite dimensional system whose
generalized coordinates are $q_{j}(t),\;$ $j=1,...,k$ are obtained as the
Euler Lagrange equations relative to the following functional
\[
\mathcal{S}=\int\mathcal{L}\left(  t,q_{1},.....,q_{k},\dot{q}_{1}%
,.....,\dot{q}_{k}\right)  dt
\]

Also the Dynamics of fields can be determined by the Variational Principle.
From a mathematical point of view a field is a function\footnote{We use the
convention to use Greek letters $\psi,\Psi$ etc. to denote complex valued
functions and Latin letters $u,v,$... to denote real valued function.}
\[
u:\mathbf{R}^{N+1}\rightarrow\mathbf{R}^{k},\;\;\;\;\;u=(u_{1},....,u_{k}).
\]
where $\mathbf{R}^{N+1}$ is the space-time continuum and $\mathbf{R}^{k}$ is
called the internal parameters space. Of course, in physical problems, the
space dimension $N$ is $1,2$ or $3.$ The space and time coordinates will be
respectively denoted by $x=\left(  x_{1},..,x_{N}\right)  $ and $t$
respectively. The function $u(t,x)$ describes the \textit{internal} state of
the ether (or vacuum) at the point $x$ and time $t.$

From a mathematical point of view, assumption \textbf{A-1} states that the
field equations are obtained by the variation of the action functional defined
as follows:
\begin{equation}
\mathcal{S}=\int\int\mathcal{L}\left(  t,x,u,\nabla u,\partial_{t}u\right)
\,dx\,dt. \label{gennifer}%
\end{equation}
The function $\mathcal{L}$ is called Lagrangian density function but in the
following, as usual, we will call it just Lagrangian function.

If $u$ is a scalar function, the variation of (\ref{gennifer}) gives the
following equation:
\begin{equation}
\sum_{i=0}^{N}\frac{\partial}{\partial x_{i}}\left(  \frac{\partial
\mathcal{L}}{\partial u_{x_{i}}}\right)  -\frac{\partial\mathcal{L}}{\partial
u}=0 \label{sei}%
\end{equation}
If $u=(u_{1},....,u_{k})$, the Euler-Lagrange equations take the same form
provided that we have use the convention that%
\[
\frac{\partial\mathcal{L}}{\partial u_{x_{i}}}=\left(  \frac{\partial
\mathcal{L}}{\partial u_{1,x_{i}}},.....,\frac{\partial\mathcal{L}}{\partial
u_{k,x_{i}}}\right)  ,\ \frac{\partial\mathcal{L}}{\partial u}=\left(
\frac{\partial\mathcal{L}}{\partial u_{1}},.....,\frac{\partial\mathcal{L}%
}{\partial u_{k}}\right)  .
\]
So, if $u$ has $k$ components ( $k>1)$ then eq. (\ref{sei}) is equivalent to
the $k$ equations:
\begin{equation}
\sum_{i=0}^{N}\frac{\partial}{\partial x_{i}}\left(  \frac{\partial
\mathcal{L}}{\partial u_{\ell,x_{i}}}\right)  -\frac{\partial\mathcal{L}%
}{\partial u_{\ell}}=0,\;\;\;\ell=1,...,k \label{seipiu}%
\end{equation}

\subsection{The invariance principle}

A functional $J$ is called invariant for a representation $T_{g}$ of a Lie
group if
\begin{equation}
J(T_{g}u)=J(u). \label{alina}%
\end{equation}
Now, let us consider the variational equation
\[
\left\{
\begin{array}
[c]{c}%
u\in X\\
F(u)=0
\end{array}
\right.
\]
where $F(u)=dJ(u).$ If $J$ is invariant, given any solution $u,$ we have that
also $T_{g}u$ is a solution.

We need to be careful in the interpretation of (\ref{alina}). In fact, if $u$
belongs to some function space $\mathfrak{F}\left(  \Omega,V\right)  $, (where
$\Omega\subset\mathbb{R}^{N+1}$ and $V$ is a finite dimensional vector space),
it might happen that $T_{g}u\notin\mathfrak{F}\left(  \Omega,V\right)  $. For
example, if%
\[
\left(  T_{h}u\right)  (x)=u(x-h),\ h\in\mathbb{R}^{N+1}%
\]
we have that $T_{h}u\in\mathfrak{F}\left(  \Omega-h,V\right)  .$

Thus we are led to give the following definition: we say the Lagrangian
$\mathcal{L}\left(  t,x,u,\nabla u,\partial_{t}u\right)  $ is invariant with
respect to the representation $T_{g}$ if%
\begin{equation}
\mathcal{L}\left(  t,x,u,\nabla u,\partial_{t}u\right)  =\mathcal{L}\left(
t^{\prime},x^{\prime},u^{\prime},\nabla u^{\prime},\partial_{t}u^{\prime
}\right)  \label{alina0}%
\end{equation}
where $u^{\prime}\left(  t^{\prime},x^{\prime}\right)  =T_{g}u\left(
t,x\right)  .$

In this case, the equation (\ref{alina}) need to be interpreted as follows:%
\begin{equation}
\int_{T_{g}\Omega}\mathcal{L}\left(  t^{\prime},x^{\prime},u^{\prime},\nabla
u^{\prime},\partial_{t}u^{\prime}\right)  dxdt=\int_{\Omega}\mathcal{L}\left(
t,x,u,\nabla u,\partial_{t}u\right)  dxdt \label{alina1}%
\end{equation}

\noindent where%
\[
T_{g}\Omega:=\left\{  \left(  t^{\prime},x^{\prime}\right)  \in\mathbb{R}%
^{N+1}:\left(  t,x\right)  \in\Omega\right\}  .
\]
We say that a Lagrangian $\mathcal{L}$ is invariant for the action $T_{g}$ if
(\ref{alina1}) holds for all bonded sets $\Omega\subset\mathbb{R}^{N}.$

\subsection{The Poincar\'{e} invariance\label{PG}}

The fundamental equations of Physics are invariant for the Poincar\'{e} group:
it is the \textit{basic} principle on which the special theory of relativity
is based.

The Poincar\'{e} group $\mathfrak{P}$ is a generalization of the isometry
group $\mathfrak{E}$. The isometry group $\mathfrak{E}$ in $\mathbb{R}^{N}$ is
the group of transformation which preserves the quadratic form%
\[
\left\vert x\right\vert ^{2}:=\sum_{i=1}^{N}x_{i}^{2}%
\]
i.e. the Euclidean norm and hence the Euclidean distance
\[
d_{E}(x,y)=\sqrt{\sum_{i=1}^{N}\left\vert x_{i}-y_{i}\right\vert ^{2}};
\]
namely, if $g\in\mathfrak{E}$,
\[
d_{E}(gx,gy)=d_{E}(x,y).
\]

If we identify the physical space with $\mathbb{R}^{3},$ the isometry group is
also called congruence group. Roughly speaking, the content of Euclidean
geometry is the study of the properties of geometric objects which are
preserved by the congruence group.

The Poincar\'{e} group $\mathfrak{P}$, by definition, is the transformation
group in $\mathbb{R}^{N+1}$ which preserves the quadratic form
\[
\left\vert x\right\vert _{M}^{2}=-x_{0}^{2}+\sum_{i=1}^{N}x_{i}^{2}%
\]
which is induced by the Minkowski bilinear form
\[
\left\langle x,y\right\rangle _{M}=-x_{0}y_{0}+\sum_{i=1}^{N}x_{i}y_{i}%
\]

The Minkowski vectors $v=\left(  v_{0},..,v_{N}\right)  \equiv\left(
v_{0,}\mathbf{v}\right)  $ are classified according to their \textit{causal
}nature as follows:

\begin{itemize}
\item a vector is called space-like if $\left\langle v,v\right\rangle _{M}>0,
$

\item a vector is called space-like if light-like if $\left\langle
v,v\right\rangle _{M}=0,$

\item a vector is called space-like if time-like if $\left\langle
v,v\right\rangle _{M}<0.$
\end{itemize}

The casual nature is not changed by a Poincar\'{e} transformation, and hence
it is not a transitive group (as the isometry group): space and time-are
mixed, but not.....so much.

In the real world we have $N=3$ and the Poincar\'{e} group is a 10 parameter
Lie group generated by the following one-parameter transformations:

\begin{itemize}
\item Space translations in the directions $x,y,z:$%
\[%
\begin{array}
[c]{c}%
x^{\prime}=x+x_{0}\\
y^{\prime}=y\\
z^{\prime}=z\\
t^{\prime}=t
\end{array}
;\;
\begin{array}
[c]{c}%
x^{\prime}=x\\
y^{\prime}=y+y_{0}\\
z^{\prime}=z\\
t^{\prime}=t
\end{array}
;\;
\begin{array}
[c]{c}%
x^{\prime}=x\\
y^{\prime}=y\\
z^{\prime}=z+z_{0}\\
t^{\prime}=t.
\end{array}
\]
This invariance guarantees that space is homogeneous, namely that the laws of
physics are independent of space: if an experiment is performed here or there,
it gives the same results.

\item Space rotations:
\[%
\begin{array}
[c]{c}%
x^{\prime}=x\\
y^{\prime}=y\cos\theta_{1}-z\sin\theta_{1}\\
z^{\prime}=y\sin\theta_{1}+z\cos\theta_{1}\\
t^{\prime}=t
\end{array}
;\;
\begin{array}
[c]{c}%
x^{\prime}=x\cos\theta_{2}-z\sin\theta_{2}\\
y^{\prime}=y\\
z^{\prime}=x\sin\theta_{2}+z\cos\theta_{2}\\
t^{\prime}=t
\end{array}
;\;
\begin{array}
[c]{c}%
x^{\prime}=x\cos\theta_{3}-y\sin\theta_{3}\\
y^{\prime}=x\sin\theta_{3}+y\cos\theta_{3}\\
z^{\prime}=z\\
t^{\prime}=t.
\end{array}
\]
This invariance guarantees that space is isotropic, namely that the laws of
physics are independent of orientation.

\item Time translations:
\[%
\begin{array}
[c]{c}%
x^{\prime}=x\\
y^{\prime}=y\\
z^{\prime}=z\\
t^{\prime}=t+t_{0}.
\end{array}
\]
This invariance guarantees that time is isotropic; namely that the laws of
physics are independent of time: if an experiment is performed earlier or
later, it gives the same results.

\item Lorentz boosts:
\begin{equation}%
\begin{array}
[c]{c}%
x^{\prime}=\gamma\left(  x-v_{1}t\right) \\
y^{\prime}=y\\
z^{\prime}=z\\
t^{\prime}=\gamma\left(  t-v_{1}x\right)
\end{array}
;\;
\begin{array}
[c]{c}%
x^{\prime}=x\\
y^{\prime}=\gamma\left(  y-v_{2}t\right) \\
z^{\prime}=z\\
t^{\prime}=\gamma\left(  t-v_{2}y\right)
\end{array}
;\;
\begin{array}
[c]{c}%
x^{\prime}=x\\
y^{\prime}=y\\
z^{\prime}=\gamma\left(  z-v_{3}t\right) \\
t^{\prime}=\gamma\left(  t-v_{3}z\right)
\end{array}
. \label{LT}%
\end{equation}

\end{itemize}

where
\begin{equation}
\gamma=\frac{1}{\sqrt{1-v^{2}}}. \label{gamma}%
\end{equation}
with $v=v_{i},\;i=1,2,3.$ This invariance is an empirical fact and, as it will
be shown in section \ref{DP}, it implies the remarkable facts of the theory of
relativity such as the space contraction, the time dilation and the equality
between mass and energy..

The Lorentz group is the 6 parameters Lie group generated by the space
rotations and the lorentz boosts (plus the time inversion, $t\rightarrow-t,$
and the parity inversion $(x,y,z)\rightarrow(-x,-y,-z)$). Clearly it is a
linear subgroup of GL(6).

The Poincar\'{e} group is the 10 parameters Lie group generated by the Lorentz
group and the space-time translations. Then it is a subgroup of the affine
group in $\mathbb{R}^{6}.$

The Poicar\`{e} group acts on a scalar field $\psi$ by the following
representation:
\[
\left(  T_{g}\psi\right)  \left(  t,x\right)  =\psi\left(  t^{\prime
},x^{\prime}\right)  ,\;\;\left(  t^{\prime},x^{\prime}\right)  =g\left(
t,x\right)
\]

The simplest equation invariant for this representation of the Poincar\'{e}
group is the D'Alembert equation:
\begin{equation}
\square\psi=0 \label{D'Alembert}%
\end{equation}
where
\[
\square\psi=\frac{\partial^{2}\psi}{\partial t^{2}}-\Delta\psi
\;\;\text{and\ \ }\Delta\psi=\frac{\partial^{2}\psi}{\partial x^{2}}%
+\frac{\partial^{2}\psi}{\partial y^{2}}+\frac{\partial^{2}\psi}{\partial
z^{2}}.
\]

The D'Alembert equation is the simplest \textit{variational }field equation
invariant for the Poincar\'{e} group.

In fact it is obtained from the variation of the action
\begin{equation}
\mathcal{S}_{0}=-\frac{1}{2}\int\left\langle d\psi,d\psi\right\rangle
_{M}\,dx\,dt=\frac{1}{2}\int\left[  \left\vert \partial_{t}\psi\right\vert
^{2}-\left\vert \nabla\psi\right\vert ^{2}\right]  \,dx\,dt. \label{semplice}%
\end{equation}
In this case, the Lagrangian function is given by
\begin{equation}
\mathcal{L}_{0}=-\frac{1}{2}\left\langle d\psi,d\psi\right\rangle _{M}%
=\frac{1}{2}\left\vert \partial_{t}\psi\right\vert ^{2}-\frac{1}{2}\left\vert
\nabla\psi\right\vert ^{2} \label{semplice2}%
\end{equation}
It is easy to check that if $\psi$ is a solution of this equation, then also
$T_{g}\psi\ $is a solutions of the equation for every $g\in\mathfrak{P}$.

\subsection{The Galileo invariance}

The Galileo group $\mathfrak{G}$ as the Poincar\'{e} group is a trasformation
group on the space time $\mathbb{R}^{4}.$ The Galileo group, by definition, if
the set of trasformations which preserves the time intevals and the Euclidean
distance between simultaneous points. More pricisely, an affine trasformation
$g\in\mathfrak{G}$ if, given two points $\left(  t_{1},x_{1}\right)  $ and
$\left(  t_{2},x_{2}\right)  ,$ we have that
\[
t_{1}^{\prime}-t_{2}^{\prime}=t_{1}-t_{2}%
\]
and
\[
t_{1}=t_{2}\Rightarrow\left(  t_{1}^{\prime}=t_{2}^{\prime}\ \ and\ \ d_{E}%
\left(  x_{1}^{\prime}-x_{2}^{\prime}\right)  =d_{E}\left(  x_{1}%
-x_{2}\right)  \right)
\]
where $\left(  t_{i}^{\prime},x_{i}^{\prime}\right)  =g\left(  t_{i}%
,x_{i}\right)  $ and $d_{E}(x,y)$ is the Euclidean distance.

Thus the Galileo group is a 10 parameters Lie group generated by the
space-time translation the space rotations but the Lorentz boosts (\ref{LT})
are replaced be the \textit{Galilean tranformations} namely by the
tranformations
\begin{equation}%
\begin{array}
[c]{c}%
x^{\prime}=x-v_{1}t\\
y^{\prime}=y\\
z^{\prime}=z\\
t^{\prime}=t
\end{array}
;\;%
\begin{array}
[c]{c}%
x^{\prime}=x\\
y^{\prime}=y-v_{2}t\\
z^{\prime}=z\\
t^{\prime}=t
\end{array}
;\;%
\begin{array}
[c]{c}%
x^{\prime}=x\\
y^{\prime}=y\\
z^{\prime}=z-v_{3}t\\
t^{\prime}=t
\end{array}
. \label{galgr}%
\end{equation}

The equations of classical mechanics are invariant for the Galileo group. We
are interested in field equations which are invariant for a representation of
the Galileo group.

Given the Galileo tranformation $g_{\mathbf{v}}:\mathbb{R}^{4}\rightarrow
\mathbb{R}^{4}$ defined by
\[
g_{\mathbf{v}}\left(  t,x\right)  =(t,x-\mathbf{v}t).
\]
we consider the representation $T_{g_{\mathbf{v}}}:L^{2}\left(  \mathbb{R}%
^{4},\mathbb{C}\right)  \rightarrow L^{2}\left(  \mathbb{R}^{4},\mathbb{C}%
\right)  $ defined by
\begin{equation}
\left(  T_{g_{\mathbf{v}}}\psi\right)  \left(  t,x\right)  =\psi\left(
t,x-\mathbf{v}t\right)  e^{i(\mathbf{v\cdot}x-\frac{1}{2}\mathbf{v}^{2}t)},
\label{gal}%
\end{equation}

The Shroedinger equation for a free particle (of mass 1)
\[
i\frac{\partial\psi}{\partial t}=-\frac{1}{2}\Delta\psi
\]
is the simplest second order equation invariant for a representation of the
Galileo group on the space of complex valued vector fields $L^{2}\left(
\mathbb{R}^{4},\mathbb{C}\right)  $.

\subsection{The Gauge invariance}

Take a function
\[
\psi:\mathbb{R}^{4}\rightarrow V
\]
and assume that on $V$ acts the representation $T_{g}$ of some group $\left(
G,\circ\right)  .$ This action induces two possible action on $\psi:$

\begin{itemize}
\item a global action: $\psi\left(  x\right)  \mapsto T_{g}\psi\left(
x\right)  $ where $g\in G$

\item a local action: $\psi\left(  x\right)  \mapsto T_{g\left(  x\right)
}\psi\left(  x\right)  $ where $g(x)\;$ is a smooth function with values in
$G.$
\end{itemize}

In the second case, we have a representation of the infinite dimensional
group
\[
\mathfrak{G}=\mathcal{C}^{\infty}\left(  \mathbb{R}^{4},G\right)
\]
equipped with the group operation
\[
\left(  g\circ h\right)  (x)=g(x)\circ h(x)
\]

If a Lagrangian $\mathcal{L}$ satisfies the following condition,
\[
\mathcal{L}(t,x,\psi,\nabla\psi,\partial_{t}\psi)=\mathcal{L}\left(
t,x,T_{g}\psi,\nabla\left(  T_{g}\psi\right)  ,\partial_{t}\left(  T_{g}%
\psi\right)  \right)  ,\;g\in G
\]
we say that it is invariant for a local action of the group $G,$ or for a
trivial gauge action of the group $G;$ if $\mathcal{L}$ satisfies the
following condition,
\[
\mathcal{L}(t,x,\psi,\nabla\psi,\partial_{t}\psi)=\mathcal{L}\left(
t,x,T_{g\left(  x\right)  }\psi,\nabla\left(  T_{g\left(  x\right)  }%
\psi\right)  ,\partial_{t}\left(  T_{g\left(  x\right)  }\psi\right)  \right)
,\;g\left(  x\right)  \in\mathfrak{G}%
\]
we say that it is invariant for a local action of the group $G,$ or for a
gauge action of the group $\mathfrak{G}.$

Let us consider two simple examples: the functional
\[
\int\mathcal{L}\left(  \nabla u\right)  \;dx\;,\;u\in\mathbb{R}%
\]
is invariant for a global action of the group $\left(  \mathbb{R}%
,\mathbb{+}\right)  .$ In fact, if we set $T_{r}u=u+r,$ $r\in\mathbb{R},$ we
have that
\[
\mathcal{L}\left(  \nabla u\right)  =\mathcal{L}\left(  \nabla\left(
T_{r}u\right)  \right)  .
\]
Next, consider the functional
\[
\int\mathcal{L}\left(  d\alpha\right)  \;dx
\]
where $\alpha$ is a 1-form and $d$ is the exterior derivative of $\alpha.$ In
this case, $\mathcal{L}\left(  d\alpha\right)  $is not only invariant for a
trivial action of $\left(  \mathbb{R},\mathbb{+}\right)  ,$ but also for the
local action
\[
T_{g(x)}\alpha=\alpha+dg(x),\;g\left(  x\right)  \in\mathfrak{G}%
:=\mathcal{C}^{\infty}\left(  \mathbb{R}^{4},\mathbb{R}\right)
\]
in fact
\[
\mathcal{L}\left(  d\left(  \alpha+dg(x)\right)  \right)  =\mathcal{L}\left(
d\alpha\right)  .
\]

The simplest gauge invariance can be obtained taking a complex valued scalar
field
\[
\psi:\mathbb{R}^{4}\rightarrow\mathbb{C},
\]
and to consider the group $S^{1}=\left\{  e^{i\theta}:\theta\in\mathbb{R}%
\right\}  $ and the following representation
\begin{equation}
\psi\mapsto e^{i\theta}\psi\label{giulia}%
\end{equation}
The Scroedinger equation and the Klein-Gordon equation are invariant for the
global action (\ref{giulia}). The Klein-Gordon-Maxwell equations are invariant
for a local action (\ref{giulia}). For a discussion of these aspects of KGM,
we refer to \cite{befogranas}.

\subsection{Noether's theorem}

In this section we will give a proof of Noether theorem stated in a suitable
form for the applications considered in this paper.

First of all, we need the following lemma

\begin{lemma}
\label{cont}Let $\mathcal{\rho}:\mathbb{R}^{N+1}\rightarrow\mathbb{R}$ and
$\mathbf{J:}\mathbb{R}^{N+1}\rightarrow\mathbb{R}^{N}$ be two smooth functions
defined on the "space-time". Assume that they satisfy the continuity equation
\begin{equation}
\frac{\partial\mathcal{\rho}}{\partial t}+\nabla\cdot\mathbf{J=}0
\label{conti}%
\end{equation}
and that for all $t$
\begin{equation}
\mathcal{\rho}(\cdot,t),\;\frac{\partial\mathcal{\rho}}{\partial t}%
(\cdot,t)\text{ and }\mathbf{J(}\cdot,t)\text{ are in }L^{1}(\mathbb{R}^{3})
\label{ass}%
\end{equation}
Then for all $t$
\begin{equation}
\frac{d}{dt}\int_{\mathbb{R}^{N}}\mathcal{\rho}(x,t)dx=0 \label{const}%
\end{equation}

\end{lemma}

\textbf{Proof. }Let
\[
B_{R}=\left\{  x\in\mathbb{R}^{3}:\left\vert x\right\vert <R\right\}  ,\text{
}R>0
\]
then, integrating on $B_{R}$, we get
\begin{equation}
\int_{B_{R}}\frac{\partial\mathcal{\rho}}{\partial t}dx=-\int_{B_{R}}%
\nabla\cdot\mathbf{J}dx=-\int_{\partial B_{R}}(\mathbf{J\cdot n)}%
d\sigma\label{div}%
\end{equation}
where $\mathbf{n}$ denotes the outward normal to the boundary $\partial B_{R}
$ of $B_{R}.$ Then
\begin{equation}
\left\vert \int_{B_{R}}\frac{\partial\mathcal{\rho}}{\partial t}dx\right\vert
\leq\int_{\partial B_{R}}\left\vert \mathbf{J\cdot n}\right\vert
d\sigma\label{imp}%
\end{equation}

Since $\frac{\partial\mathcal{\rho}}{\partial t}(.,t)$ is in $L^{1}%
(\mathbb{R}^{3}),$ there exists $\underset{R\rightarrow\infty}{\lim}\left\vert
\int_{B_{R}}\frac{\partial\mathcal{\rho}}{\partial t}dx\right\vert ,$ and we
have to prove that this limit is $0.$ Arguing by contradiction we assume that
\begin{equation}
\underset{R\rightarrow\infty}{\lim}\left\vert \int_{B_{R}}\frac{\partial
\mathcal{\rho}}{\partial t}dx\right\vert =\alpha>0 \label{impo}%
\end{equation}

then, by (\ref{imp}) and (\ref{impo}), the map $\varphi$ defined by
\[
\varphi(R)=\int_{\partial B_{R}}\left\vert \mathbf{J\cdot n}\right\vert
d\sigma
\]
is not integrable in $(0,+\infty)$ and
\[
\int_{\mathbb{R}^{N}}\left\vert \mathbf{J\cdot n}\right\vert dx=\int
_{0}^{+\infty}\varphi(R)dR=+\infty
\]
which contradicts assumption (\ref{ass}).

$\square$\bigskip

Suppose that a Lagrangian is invariant for the action $T_{g}$ of some Lie
group $G.$ We denote by $T_{g\left(  \lambda\right)  }$ ($\lambda\in
\mathbb{R}$) the action of a one-parameter subgroup $\left\{  g\left(
\lambda\right)  \right\}  _{\lambda\in\mathbb{R}}.$ Notice that this subgroup
is isomorphic either to $S^{1}$ or $\mathbb{R}.$ We use the following
notation:%
\begin{equation}
u_{\lambda}=T_{g\left(  \lambda\right)  }u \label{pi1}%
\end{equation}
and, if the group acts also on the independent variables, we set%
\begin{align}
t_{\lambda}  &  =T_{g\left(  \lambda\right)  }t\label{pi2}\\
x_{\lambda}  &  =T_{g\left(  \lambda\right)  }x \label{pi3}%
\end{align}

For example, consider the first of the Lorentz transformation (\ref{LT}); in
this case the parameter $\lambda$ is the first component of the velocity $v, $
and we have%
\begin{align*}
t_{v}  &  =\frac{t-vx}{\sqrt{1-v^{2}}}\\
x_{1,v}  &  =\frac{x_{1}-vt}{\sqrt{1-v^{2}}}\\
u_{v}  &  =u\left(  t_{v},x_{1,v},x_{2},x_{3}\right)
\end{align*}

\begin{lemma}
\label{elle}If $\mathcal{L}$ is invariant with respect to a one parameter
group $g\left(  \lambda\right)  ,$ then, using the notation (\ref{pi1}%
),(\ref{pi2}),(\ref{pi3}), we have that%
\[
\left.  \frac{d}{d\lambda}\right\vert _{\lambda=0}\int_{\Omega}\mathcal{L}%
\left(  t_{\lambda},x_{\lambda},u_{\lambda},\nabla u_{\lambda},\partial
_{t}u_{\lambda}\right)  \varphi\left(  t_{\lambda},x_{\lambda}\right)
\,dxdt=0
\]
where $\Omega=\left[  t_{0},t_{1}\right]  \times\mathbb{R}^{N}$ and
$\varphi\in\mathcal{D}\left(  \Omega\right)  $\footnote{$\mathcal{D}\left(
\Omega\right)  $ denotes the space of infinitely differentiable functions with
compact support in $\Omega.$}.
\end{lemma}

\textbf{Proof.} We approximate $\varphi$ with a function $\varphi
_{\varepsilon}$ defined by%
\[
\varphi_{\varepsilon}=\sum_{j}j\varepsilon\chi_{\Omega_{j}}%
\]
where $\chi_{\Omega_{j}}$ is the characteristic function of%
\[
\Omega_{j}:=\left\{  (t,x)\in\mathbb{R}^{N+1}:j\varepsilon\leq\varphi
(t,x)<\left(  j+1\right)  \varepsilon\right\}
\]
If $\lambda$ is small, so that the support of $\varphi\left(  t_{\lambda
},x_{\lambda}\right)  $ is contained in $\Omega,$ we have that
\[
\int_{\Omega}\mathcal{L}\left(  t_{\lambda},x_{\lambda},u_{\lambda},\nabla
u_{\lambda},\partial_{t}u_{\lambda}\right)  \varphi_{\varepsilon}\left(
t_{\lambda},x_{\lambda}\right)  \,dxdt=\sum_{j}j\varepsilon\int_{T_{\lambda
}\Omega_{j}}\mathcal{L}\left(  t_{\lambda},x_{\lambda},u_{\lambda},\nabla
u_{\lambda},\partial_{t}u_{\lambda}\right)  \,dxdt
\]
and hence, by (\ref{alina1}),%
\begin{align*}
\sum_{j}j\varepsilon\int_{T_{\lambda}\Omega_{j}}\mathcal{L}\left(  t_{\lambda
},x_{\lambda},u_{\lambda},\nabla u_{\lambda},\partial_{t}u_{\lambda}\right)
\,dxdt  &  =\sum_{j}j\varepsilon\int_{\Omega_{j}}\mathcal{L}\left(
t,x,u,\nabla u,\partial_{t}u\right)  \,dxdt\\
&  =\int_{\Omega}\mathcal{L}\left(  t,x,u,\nabla u,\partial_{t}u\right)
\varphi_{\varepsilon}\left(  t,x\right)  \,dxdt
\end{align*}
and so,%
\[
\int_{\Omega}\mathcal{L}\left(  t_{\lambda},x_{\lambda},u_{\lambda},\nabla
u_{\lambda},\partial_{t}u_{\lambda}\right)  \varphi_{\varepsilon}\left(
t_{\lambda},x_{\lambda}\right)  \,dxdt=\int_{\Omega}\mathcal{L}\left(
t,x,u,\nabla u,\partial_{t}u\right)  \varphi\left(  t,x\right)  \,dxdt
\]
Taking the limit for $\varepsilon\rightarrow0,$ we get that%
\[
\int_{\Omega}\mathcal{L}\left(  t_{\lambda},x_{\lambda},u_{\lambda},\nabla
u_{\lambda},\partial_{t}u_{\lambda}\right)  \varphi\left(  t_{\lambda
},x_{\lambda}\right)  \,dxdt=\int_{\Omega}\mathcal{L}\left(  t,x,u,\nabla
u,\partial_{t}u\right)  \varphi\left(  t,x\right)  \,dxdt
\]
and we get the conclusion.

$\square$

\bigskip

The above lemma is useful since the introduction of the compact support
function $\varphi$ allows us to work on a fixed domain $\Omega$ and hence we
do not have to consider the variation of the domain $T_{\lambda}\Omega.$

\begin{theorem}
\label{noe} Let $\mathcal{L}$ be invariant with respect to a one parameter
group $g\left(  \lambda\right)  ,$ and let $u=u_{\lambda}$ be a smooth
solution ot the Euler-Lagrange equation (\ref{sei}). Using the notation
(\ref{pi1}),(\ref{pi2}), (\ref{pi3}), we set
\begin{equation}
\rho=\left(  \frac{\partial\mathcal{L}}{\partial u_{\lambda,t}}\frac{\partial
u_{\lambda}}{\partial\lambda}-\mathcal{L}\frac{\partial t_{\lambda}}%
{\partial\lambda}\right)  _{\lambda=0} \label{bellissima}%
\end{equation}

\noindent and%
\begin{equation}
\mathbf{J}=\sum_{i=1}^{N}\left(  \frac{\partial\mathcal{L}}{\partial
u_{\lambda,x^{i}}}\frac{\partial u_{\lambda}}{\partial\lambda}-\mathcal{L}%
\frac{\partial x_{\lambda}^{i}}{\partial\lambda}\right)  _{\lambda
=0}\mathbf{e}_{i}. \label{carino}%
\end{equation}
Then we get the continuity equation
\begin{equation}
\frac{\partial\rho}{\partial t}+\nabla\cdot\mathbf{J=}0. \label{ec}%
\end{equation}

\end{theorem}

\textbf{Proof. }By the invariance of the Lagrangian and lemma \ref{elle}, we have:%

\[
\frac{d}{d\lambda}\int\mathcal{L}\varphi\,dxdt=0\;\;\forall\varphi
\in\mathcal{D}\left(  \Omega\right)
\]
and hence
\begin{equation}
\int\left(  \frac{d\mathcal{L}}{d\lambda}\varphi+\mathcal{L}\frac{d\varphi
}{d\lambda}\right)  \,dxdt=0\;\;\forall\varphi\in\mathcal{D}\left(
\Omega\right)  \label{nina}%
\end{equation}
Let us compute each derivative; in this computation, we write $u,x,t$ instead
of $u_{\lambda},x_{\lambda},t_{\lambda}$ to make it readable and we use the
notation $x^{0}=t$:%
\begin{align}
\frac{d\mathcal{L}}{d\lambda}  &  =\sum_{i=0}^{N}\frac{\partial\mathcal{L}%
}{\partial u_{x^{i}}}\frac{\partial^{2}u}{\partial\lambda\partial x^{i}}%
+\frac{\partial\mathcal{L}}{\partial u}\frac{\partial u}{\partial\lambda
}\nonumber\\
&  =\sum_{i=0}^{N}\left[  \frac{\partial}{\partial x^{i}}\left(
\frac{\partial\mathcal{L}}{\partial u_{x^{i}}}\frac{\partial u}{\partial
\lambda}\right)  -\frac{\partial}{\partial x^{i}}\left(  \frac{\partial
\mathcal{L}}{\partial u_{x^{i}}}\right)  \frac{\partial u}{\partial\lambda
}\right]  +\frac{\partial\mathcal{L}}{\partial u}\frac{\partial u}%
{\partial\lambda}\nonumber\\
&  =\sum_{i=0}^{N}\frac{\partial}{\partial x^{i}}\left(  \frac{\partial
\mathcal{L}}{\partial u_{x^{i}}}\frac{\partial u}{\partial\lambda}\right)
-\left[  \sum_{i=0}^{N}\frac{\partial}{\partial x^{i}}\left(  \frac
{\partial\mathcal{L}}{\partial u_{x^{i}}}\right)  -\frac{\partial\mathcal{L}%
}{\partial u}\right]  \frac{\partial u}{\partial\lambda}\nonumber
\end{align}
Then, by equation (\ref{sei})%
\begin{equation}
\frac{d\mathcal{L}}{d\lambda}=\sum_{i=0}^{N}\frac{\partial}{\partial x^{i}%
}\left(  \frac{\partial\mathcal{L}}{\partial u_{x^{i}}}\frac{\partial
u}{\partial\lambda}\right)  . \label{gina}%
\end{equation}

Also
\[
\frac{d\varphi}{d\lambda}=\sum_{i=0}^{N}\frac{\partial\varphi}{\partial x^{i}%
}\frac{\partial x^{i}}{\partial\lambda}%
\]
Then, by (\ref{nina}), (\ref{gina}) and the above equality, we have that%
\begin{equation}
\int\left[  \sum_{i=0}^{N}\frac{\partial}{\partial x^{i}}\left(
\frac{\partial\mathcal{L}}{\partial u_{x^{i}}}\frac{\partial u}{\partial
\lambda}\right)  \varphi+\mathcal{L}\sum_{i=0}^{N}\frac{\partial\varphi
}{\partial x^{i}}\frac{\partial x^{i}}{\partial\lambda}\right]  \,dxdt=0
\label{ciucca}%
\end{equation}
Moreover, since $\varphi$ has compact support, by the divergence theorem we
have that%
\[
\int\sum_{i=0}^{N}\frac{\partial}{\partial x^{i}}\left(  \mathcal{L}%
\frac{\partial x^{i}}{\partial\lambda}\varphi\right)  \,dxdt=0;
\]
so
\[
\int\sum_{i=0}^{N}\frac{\partial}{\partial x^{i}}\left(  \mathcal{L}%
\frac{\partial x^{i}}{\partial\lambda}\right)  \varphi\,dxdt+\int
\mathcal{L}\sum_{i=0}^{N}\frac{\partial\varphi}{\partial x^{i}}\frac{\partial
x^{i}}{\partial\lambda}\,dxdt=0
\]
By this equality and (\ref{ciucca}), we get that
\begin{align*}
0  &  =\int\sum_{i=0}^{N}\frac{\partial}{\partial x^{i}}\left(  \frac
{\partial\mathcal{L}}{\partial u_{x^{i}}}\frac{\partial u}{\partial\lambda
}\right)  \varphi\ dxdt-\int\sum_{i=0}^{N}\frac{\partial}{\partial x^{i}%
}\left(  \mathcal{L}\frac{\partial x^{i}}{\partial\lambda}\right)
\varphi\,dxdt\\
&  =\int\left[  \sum_{i=0}^{N}\frac{\partial}{\partial x^{i}}\left(
\frac{\partial\mathcal{L}}{\partial u_{x^{i}}}\frac{\partial u}{\partial
\lambda}-\mathcal{L}\frac{\partial x^{i}}{\partial\lambda}\right)
\varphi\right]  \,dxdt
\end{align*}

By the arbitrariness of $\varphi$ we get
\[
\sum_{i=0}^{N}\frac{\partial}{\partial x^{i}}\left(  \frac{\partial
\mathcal{L}}{\partial u_{x^{i}}}\frac{\partial u}{\partial\lambda}%
-\mathcal{L}\frac{\partial x^{i}}{\partial\lambda}\right)  =0
\]
or,
\[
\frac{\partial}{\partial t}\left(  \frac{\partial\mathcal{L}}{\partial u_{t}%
}\frac{\partial u}{\partial\lambda}-\mathcal{L}\frac{\partial t}%
{\partial\lambda}\right)  +\sum_{i=1}^{N}\frac{\partial}{\partial x^{i}%
}\left(  \frac{\partial\mathcal{L}}{\partial u_{x^{i}}}\frac{\partial
u}{\partial\lambda}-\mathcal{L}\frac{\partial x^{i}}{\partial\lambda}\right)
\]

Then the functions (\ref{bellissima}) and (\ref{carino}) satisfy the
continuity equation (\ref{ec}).

$\square$

\bigskip

Then by lemma \ref{cont} and Th. \ref{noe}, we have the following result:

\begin{theorem}
\label{noether}(\textbf{Noether's theoerem}) Let $\mathcal{L}$ be invariant
with respect to a one parameter group $g\left(  \lambda\right)  ,$ and let $u$
be a smooth solution of the Euler-Lagrange equation (\ref{sei}). Suppose that
$u$ decays sufficiently fast so that (\ref{ass}) holds. Then, using the
notation (\ref{pi1}),(\ref{pi2}),(\ref{pi3}), we have that
\[
\mathcal{I}\left(  u\right)  =\int\left(  \frac{\partial\mathcal{L}}{\partial
u_{\lambda,t}}\frac{\partial u_{\lambda}}{\partial\lambda}-\mathcal{L}%
\frac{\partial t_{\lambda}}{\partial\lambda}\right)  _{\lambda=0}dx
\]
is an integral of motion.
\end{theorem}

\subsection{Conservation laws\label{cl}}

Now, using Noether theorem \ref{noether}, we can compute the main integral of
motion. They are due to the homogeneity of time and the homogeneity and
isotropy of space which provide the invariance with respect to the time
translations, space translations and space rotations. We consider the case in
which $\mathcal{L}$ depends on a complex valued scalar function $\psi.$ The
computation can be done setting $\psi=u_{1}+iu_{2},$ and considering
$\mathcal{L}$ as function of $u=\left(  u_{1},u_{2}\right)  .$

\begin{itemize}
\item \textbf{Energy}. Energy, by definition, is the quantity which is
preserved by the time invariance of the Lagrangian; it has the following form%
\begin{equation}
\mathcal{E}=\operatorname{Re}\int\left(  \frac{\partial\mathcal{L}}%
{\partial\left(  \partial_{t}\psi\right)  }\cdot\overline{\partial_{t}\psi
}-\mathcal{L}\right)  dx \label{ener}%
\end{equation}

\item \textbf{Momentum. }Momentum, by definition, is the quantity which is
preserved by the space invariance of the Lagrangian; the invariance for
translations in the $x_{i}$ direction gives the following invariant:%
\[
P_{i}=-\operatorname{Re}\int\frac{\partial\mathcal{L}}{\partial\left(
\partial_{t}\psi\right)  }\cdot\overline{\partial_{x_{i}}\psi}\,dx
\]
The numbers $P_{i}$ are the components of the vector
\begin{equation}
\mathbf{P}=-\operatorname{Re}\int\frac{\partial\mathcal{L}}{\partial\left(
\partial_{t}\psi\right)  }\cdot\overline{\nabla\psi}\,dx \label{mom}%
\end{equation}

\item \textbf{Angular momentum. }The angular momentum, by definition, is the
quantity which is preserved by virtue of the invariance under space rotations
of the Lagrangian with respect to the origin%
\[
\mathbf{M}=\operatorname{Re}\int\frac{\partial\mathcal{L}}{\partial\left(
\partial_{t}\psi\right)  }\cdot\overline{\left(  \mathbf{x}\times\nabla
\psi\right)  }\;dx.
\]

\end{itemize}

\textbf{Proof. }First, we compute\textbf{\ }$M_{3}.$ Setting $\mathbf{x}%
=(x,y,z),$ we have that%
\begin{align*}
x_{\lambda}  &  =x\cos\lambda-y\sin\lambda\\
y_{\lambda}  &  =x\sin\lambda+y\cos\lambda\\
z_{\lambda}  &  =z\\
t_{\lambda}  &  =t
\end{align*}
then, setting $\psi=u_{1}+iu_{2},$
\begin{align*}
\left(  \frac{\partial\mathcal{L}}{\partial u_{1,t}}\frac{\partial u_{1,t}%
}{\partial\lambda}+\frac{\partial\mathcal{L}}{\partial u_{2,t}}\frac{\partial
u_{2,t}}{\partial\lambda}-\mathcal{L}\frac{\partial t}{\partial\lambda
}\right)  _{\lambda=0}  &  =\operatorname{Re}\left(  \frac{\partial
\mathcal{L}}{\partial\psi_{t}}\overline{\frac{\partial\psi}{\partial\lambda}%
}-\mathcal{L}\frac{\partial t}{\partial\lambda}\right)  _{\lambda=0}\\
&  =\operatorname{Re}\left[  \frac{\partial\mathcal{L}}{\partial\left(
\partial_{t}\psi\right)  }\overline{\left(  -\frac{\partial\psi}{\partial
x}y+\frac{\partial\psi}{\partial y}x\right)  }\right]
\end{align*}
Analogously, we have%
\begin{align*}
M_{1}  &  =\operatorname{Re}\int\frac{\partial\mathcal{L}}{\partial\left(
\partial_{t}\psi\right)  }\cdot\overline{\left(  \frac{\partial\psi}{\partial
z}y-\frac{\partial\psi}{\partial y}z\right)  }\\
M_{2}  &  =\operatorname{Re}\int\frac{\partial\mathcal{L}}{\partial\left(
\partial_{t}\psi\right)  }\cdot\overline{\left(  \frac{\partial\psi}{\partial
x}z-\frac{\partial\psi}{\partial z}x\right)  }%
\end{align*}
Then we get the conclusion.

$\square$

\bigskip

\textbf{Hylenic charge. }The hylenic charge, by definition, is the quantity
which is preserved by by the trivial gauge action (\ref{giulia}). The charge
has the following expression
\begin{equation}
\mathcal{H}=\operatorname{Im}\int\frac{\partial\mathcal{L}}{\partial
(\partial_{t}\psi)}\cdot\overline{\psi}\;dx\;\; \label{hylo}%
\end{equation}
\textbf{Proof.} We have that $\psi_{\lambda}=\psi e^{i\lambda};$ then%
\begin{align*}
\operatorname{Re}\left(  \frac{\partial\mathcal{L}}{\partial\left(
\psi_{\lambda}\right)  _{t}}\overline{\frac{\partial\psi_{\lambda}}%
{\partial\lambda}}-\mathcal{L}\frac{\partial t}{\partial\lambda}\right)
_{\lambda=0}  &  =\operatorname{Re}\left(  \frac{\partial\mathcal{L}}%
{\partial(\partial_{t}\psi)}i\overline{\psi}\right) \\
&  =\operatorname{Im}\left(  \frac{\partial\mathcal{L}}{\partial(\partial
_{t}\psi)}\overline{\psi}\right)
\end{align*}

$\square$

\subsection{The Hamilton-Jacobi theory}

In order to understand the motion of hylomorphic solitary waves, it is
necessary to know the basic notions of the Hamilton-Jacobi formulation of the
laws of Mechanics. Then, in this section, we will briefly recall these notions.

The Lagrangian formulation of the laws of the Mechanics assumes a function
\[
\mathcal{L}=\mathcal{L}\left(  t,q,\dot{q}\right)
\]
of the generalized coordinates of the system $q=(q_{1},...,q_{k}),$ of their
derivatives $\dot{q}=(\dot{q}_{1},...,\dot{q}_{k})$ and of time. The
trajectories $q(t)$ such that $q\left(  t_{0}\right)  =x_{0}$ and $q\left(
t_{1}\right)  =x_{1}$ are the critical points of the action functional
\begin{equation}
\mathcal{S}\left(  q\right)  =\int_{t_{0}}^{t_{1}}\mathcal{L}\left(
t,q,\dot{q}\right)  \;dt \label{azi}%
\end{equation}
defined on the space
\[
\mathcal{C}_{x_{0},x_{1}}^{1}\left[  t_{0},t_{1}\right]  =\left\{
q\in\mathcal{C}^{1}\left[  t_{0},t_{1}\right]  :q\left(  t_{0}\right)
=x_{0}\;\text{and }q\left(  t_{1}\right)  =x_{1}\right\}
\]

Thus a trajectory $q(t)$ satisfies the \textquotedblright
Euler-Lagrange\textquotedblright\ equations:
\begin{equation}
\frac{d}{dt}\frac{\partial\mathcal{L}}{\partial\dot{q}_{j}}-\frac
{\partial\mathcal{L}}{\partial q_{j}}=0,\ \ j=1,...,k \label{eula}%
\end{equation}

However, this is not the only formulation of the lows of Mechanics. An other
very important formulation can be obtained as a first order system provided
that
\begin{equation}
\frac{\partial^{2}\mathcal{L}}{\partial\dot{q}_{j}^{2}}>0. \label{lpos}%
\end{equation}
In this case, we set
\begin{equation}
p_{j}=\frac{\partial\mathcal{L}}{\partial\dot{q}_{j}}\left(  t,q,\dot
{q}\right)  \label{pi0}%
\end{equation}
By (\ref{lpos}), we have that the function
\[
\dot{q}_{j}\mapsto\frac{\partial\mathcal{L}}{\partial\dot{q}_{j}}\left(
t,q,\dot{q}\right)
\]
is smoothly invertible and hence there exits a smooth function $F$ such that
(\ref{pi0}) can be rewritten as follows:
\begin{equation}
\dot{q}=F\left(  t,q,p\right)  \label{qpunto}%
\end{equation}
Now, we can define the \textit{Hamiltonian function} as follows:
\[
\mathcal{H}\left(  t,q,p\right)  =\left[  \left\langle p,\dot{q}\right\rangle
-\mathcal{L}\left(  q,\dot{q},t\right)  \right]  _{\dot{q}=F\left(  p\right)
}%
\]
where
\[
\left\langle p,\dot{q}\right\rangle =\sum_{j=1}^{k}p_{j}\dot{q}_{j}%
\]
denotes the paring between the tangent space (to the space of the $q$'s) and
the relative cotangent space.

Then, the action (\ref{azi}) can be rewritten as follows
\[
\mathcal{S}\left(  p,q\right)  =\int_{t_{0}}^{t_{1}}\left[  \left\langle
p,\dot{q}\right\rangle -\mathcal{H}\left(  t,q,p\right)  \right]  \;dt
\]
and the relative \textquotedblright Euler-Lagrange\textquotedblright%
\ equations take the form
\begin{align*}
\dot{q}  &  =\frac{\partial\mathcal{H}}{\partial p}\left(  t,q,p\right) \\
\dot{p}  &  =-\frac{\partial\mathcal{H}}{\partial q}\left(  t,q,p\right)
\end{align*}
This is the Hamiltonian formulation of the laws of Mechanics and the above
equations are called Hamilton equations. An other equivalent formulation of
the laws of Dynamics is given by the \textit{Hamilton-Jacobi} theory which
uses notions both of the Lagrangian formulation and the Hamiltonian one. It
reduces the laws of Mechanics to a partial differential equation and to a
first order ordinary differential equation. The Hamilton-Jacobi theory has
been very useful to relate the laws of Optics to Dynamics. For us, it
essential if we want to understand the motion of solitons regarded as material particles.

The starting point is the definition of a function $S=S(t,x)$ called action.
We fix once for ever a point $\left(  t_{0},x_{0}\right)  $ and a point
$(t,x)$ which will be considered variable. Moreover, we set
\begin{equation}
S(t,x)=\int_{t_{0}}^{t}\mathcal{L}\left(  t,q_{x},\dot{q}_{x}\right)  \;dt
\label{sazi}%
\end{equation}
where $q_{x}$ is a critical point of (\ref{azi}) on the space $\mathcal{C}%
_{x_{0},x}^{1}\left[  t_{0},t\right]  ;$ in general, this point is not unique;
however, if (\ref{lpos}) holds, it is possible to prove the uniqueness of the
minimum provided that $(t,x)$ is sufficiently close to $\left(  t_{0}%
,x_{0}\right)  $ $(t\neq t_{0}).$ Hence, there exists an open set $\Omega$ in
which the function (\ref{sazi}) is well defined. The function (\ref{sazi}) is
called \textit{action }as the functional (\ref{azi}). However, even if they
have formal similar definitions, they are quite different objects: $S$ is a
functions of $k+1$ variables defined in an open set $\Omega\subset
\mathbb{R}^{N+1}$ while $\mathcal{S}$ is a functional defined in the function
space $\mathcal{C}_{x_{0},x}^{1}\left[  t_{0},t\right]  $.

The Hamilton-Jacobi theory states that the function $S$, in $\Omega,$
satisfies the following partial differential equation:
\begin{equation}
\partial_{t}S+\mathcal{H}\left(  t,x,\nabla S\right)  =0 \label{haj}%
\end{equation}

Moreover, this result can be inverted in the sense stated by the following theorem:

\begin{theorem}
Let $S$ be a solution of eq. (\ref{haj}) in $\Omega$ and let $q=(q_{1}%
,...,q_{k})$ be a solution of the following Cauchy problem for $(t,x)\in
\Omega$:
\begin{align}
\frac{\partial\mathcal{L}}{\partial\dot{q}_{j}}(t,q_{j},\dot{q}_{j})  &
=\nabla S(t,q_{j}),\ j=1,...,k\label{eleonora}\\
q\left(  \bar{t}\right)   &  =\bar{x}\nonumber
\end{align}
with $\left(  \bar{t},\bar{x}\right)  \in\Omega.$ Then, $q$ satisfies eq.
(\ref{eula}) with initial conditions
\begin{align*}
q\left(  \bar{t}\right)   &  =\bar{x}\\
\dot{q}\left(  \bar{t}\right)   &  =F\left(  \bar{t},\bar{x},\nabla S\left(
\bar{t},\bar{x}\right)  \right)
\end{align*}
where $F$ is given by (\ref{qpunto}).
\end{theorem}

The proof of this theorem can be found in any book of Classical Mechanics; for
example in the beautiful book of Landau and Lifchitz \cite{landaumeq}.

Notice that the above Cauchy problem is well posed, at least for small times,
since, by (\ref{qpunto}), eq. (\ref{eleonora}) gets the form
\[
\dot{q}=F\left(  t,q,\nabla S(t,q)\right)
\]
where $F$ is the smooth function given by (\ref{qpunto}).

Thus, we can say that the equation of motions (\ref{eula}) are equivalent to
the set of equations,
\begin{equation}
\partial_{t}S+\mathcal{H}\left(  t,x,\nabla S\right)  =0 \label{tina1}%
\end{equation}%
\begin{equation}
\dot{q}=F\left(  t,q,\nabla S(t,q)\right)  \label{tina2}%
\end{equation}

If $\mathcal{L}$ does not depend on $t,$ then $\mathcal{H}$ is a constant of
motion (namely it is the energy of the system). In this case, by eq.
(\ref{haj}), $\partial_{t}S=-h$, namely it does not depend on time and it
represents the energy of the system with the sign changed. In this case, eq.
(\ref{haj}) takes the form
\[
\mathcal{H}\left(  x,\nabla S\right)  =h.
\]
Let see some examples:

\textbf{Newtonian dynamics:}
\[
\mathcal{L}\left(  t,q,\dot{q}\right)  =\frac{1}{2}m\dot{q}^{2}-V(q)
\]
Then
\[
p=m\dot{q}%
\]%
\[
\mathcal{H}=\frac{1}{2m}p^{2}+V(q)
\]
and equations (\ref{tina1},\ref{tina2}) take the form
\begin{equation}
\partial_{t}S+\frac{1}{2m}\left\vert \nabla S\right\vert ^{2}+V(x)=0
\label{hjN}%
\end{equation}%
\begin{equation}
\dot{q}=\frac{1}{m}\nabla S(t,q) \label{hjN2}%
\end{equation}

\textbf{Relativistic dynamics: }The Lagrangian of a relativistic particle is
given by:
\[
\mathcal{L}\left(  t,q,\dot{q}\right)  =-m_{0}\sqrt{1-\dot{q}^{2}}%
\]
where $m_{0}$ is a parameter. We refer to Landau-Lifchitz \cite{landau} for a
very elegant deduction of this Lagrangian from the Mikowsky geometry of space-time.

Then
\begin{equation}
p=\frac{\partial\mathcal{L}}{\partial\dot{q}}=\frac{m_{0}}{\sqrt{1-\dot{q}%
^{2}}}\ \dot{q}=\gamma m_{0}\dot{q} \label{prel}%
\end{equation}
with%
\[
\gamma=\frac{1}{\sqrt{1-\dot{q}^{2}}}%
\]
and equation (\ref{qpunto}) becomes
\begin{equation}
\dot{q}=\frac{p}{\sqrt{m_{0}^{2}+p^{2}}} \label{qpunt1}%
\end{equation}

Since the mass of a particle is defined by the equation $m=p/\dot{q},\;$we
will get that the mass changes with velocity
\[
m=\gamma m_{0}%
\]
and the interpretation of $m_{0}$ as rest mass. The Hamiltonian is:
\[
\mathcal{H}=p\dot{q}+m_{0}\sqrt{1-\dot{q}^{2}}=\frac{m_{0}}{\sqrt{1-\dot
{q}^{2}}}\dot{q}^{2}+m_{0}\sqrt{1-\dot{q}^{2}}=\frac{m_{0}}{\sqrt{1-\dot
{q}^{2}}}=\gamma m_{0}%
\]
Since the Lagrangian in independent of time, the Hamiltonian represents the
energy and this gives the Einstein equation:
\begin{equation}
\mathcal{E}=\mathcal{H}=m=\gamma m_{0} \label{erel}%
\end{equation}

Now let express $\mathcal{H}$ as function of $p$. Using eq. (\ref{qpunt1}) we
get
\begin{equation}
\mathcal{H}\left(  p,q\right)  =\frac{m_{0}}{\sqrt{1-\frac{p^{2}}{p^{2}%
+m_{0}^{2}}}}=\sqrt{m_{0}^{2}+p^{2}} \label{hamrel}%
\end{equation}
and the equations (\ref{tina1},\ref{tina2}) take the form
\begin{equation}
\partial_{t}S+\sqrt{m_{0}^{2}+\left\vert \nabla S\right\vert ^{2}}=0
\label{tinarel1}%
\end{equation}%
\begin{equation}
\dot{q}=\frac{\nabla S}{\sqrt{m_{0}^{2}+\left\vert \nabla S\right\vert ^{2}}}
\label{tinarel2}%
\end{equation}

\section{Hylomorphic solitary waves and solitons\label{HS}}

\subsection{An abstract definition of solitary waves and solitons\label{be}}

Solitary waves and solitons are particular \textit{states} of a dynamical
system described by one or more partial differential equations. Thus, we
assume that the states of this system are described by one or more
\textit{fields} which mathematically are represented by functions
\begin{equation}
\Psi:\mathbb{R}^{N}\rightarrow V \label{lilla}%
\end{equation}
where $V$ is a vector space with norm $\left\vert \ \cdot\ \right\vert _{V}$
which is called the internal parameters space. We assume the system to be
deterministic; this means that it can be described as a dynamical system
$\left(  X,U\right)  $ where $X$ is the set of the states and $U:\mathbb{R}%
\times X\rightarrow X$ is the time evolution map. If $\Psi_{0}(x)\in X,$ the
evolution of the system will be described by the function
\begin{equation}
\Psi\left(  t,x\right)  =U_{t}\Psi_{0}(x) \label{flusso}%
\end{equation}
Now we can give a formal definition of solitary wave:

\begin{definition}
\label{solw}A state $\Psi_{0}\in X,$ is called solitary wave if its
evolution$\ $has the following form:
\[
U_{t}\Psi_{0}(x)=h_{t}\Psi_{0}(g_{t}x)
\]
where $h_{t}$ and $g_{t}$ are transformations of $V$ and $\mathbb{R}^{N}$ respectively.
\end{definition}

For example, consider a solution of a field equation which has the following
form
\[
\Psi\left(  t,x\right)  =e^{-i(\mathbf{k\cdot x-}\omega t)}\Psi_{0}%
(x-vt);\ \Psi_{0}\in L^{2}(\mathbb{R}^{N},\mathbb{C});
\]
then the conditions of the above definition are satisfied with $h_{t}%
\psi=e^{-i(\mathbf{k\cdot x-}\omega t)}\psi$ ($\psi\in\mathbb{C}$) and
$g_{t}x=x-vt.$

The solitons are solitary waves characterized by some form of stability. To
define them at this level of abstractness, we need to recall some well known
notion in the theory of dynamical systems.

\begin{definition}
Let $X$ be a metric space and let $\left(  X,U\right)  $ be a dynamical
system. An invariant set $\Gamma\subset X$ is called stable, if $\forall
\varepsilon>0,$ $\exists\delta>0,\;\forall\Psi\in X$,
\[
d(\Psi,\Gamma)\leq\delta,
\]
implies that
\[
\forall t\in\mathbb{R},\;\;d(U_{t}\Psi,\Gamma)\leq\varepsilon
\]

\end{definition}

\bigskip

\begin{definition}
\label{dos}A state $\Psi_{0}$ is called orbitally stable if there exists a
finite dimensional invariant \emph{stable} manifold $\Gamma$ such that
$\Psi_{0}\in\Gamma.$
\end{definition}

The above definition needs some explanation. Since $\Gamma$ is invariant,
$U_{t}\Psi_{0}\in\Gamma$ for every time. Thus, since $\Gamma$ is finite
dimensional, the evolution of $\Psi_{0}$ is described by a finite number of
parameters$.$ Thus the dynamical system $\left(  \Gamma,U\right)  $\ behaves
as a point in a finite dimensional phase space. By the stability of $\Gamma$,
a small perturbation of $\Psi_{0}$ remains close to $\Gamma.$ However, in this
case, its evolution depends on an infinite number of parameters. Then, as time
goes on, the evolution of the perturbed system might become very different
from $U_{t}\Psi_{0}$. Thus, this system appears as a finite dimensional system
with a small perturbation. We refer to section (\ref{dynshr}) where this fact
will be seen in details in a concrete case.

\begin{definition}
\label{ds}A state $\Psi_{0}\in X,\;\Psi_{0}\neq0,\;$ is called soliton if it
is a orbitally stable solitary wave.
\end{definition}

In general, $\dim\left(  \Gamma\right)  >N$ and hence, the \textquotedblright
state\textquotedblright\ of a soliton is described by $N$ parameters which
define its position and other parameters which define its \textquotedblright
internal state\textquotedblright.

\subsection{Definition of hylomorphic solitons}

Now let us assume that our system satisfies assumptions \textbf{A-1,}
\textbf{A-2 }and \textbf{A-3} of pag.\pageref{assA}. By Noether theorem,
assumptions \textbf{A-1} and \textbf{A-2} of guarantee the conservation of the
energy $\mathcal{E}\left(  \Psi\right)  $ (see (\ref{ener})) while
\textbf{A-1} and \textbf{A-3 }guarantee the conservation ot the
\textit{hylenic charge} $\mathcal{H}\left(  \Psi\right)  $ (see (\ref{hylo})).

\textit{\ }

\begin{definition}
\label{hys}A stationary wave $\Psi_{0}$ is called hylomorphic wave if
\begin{equation}
\mathcal{E}\left(  \Psi_{0}\right)  =c_{\sigma}:=\underset{\Psi\in
\mathfrak{M}_{\sigma}}{\min}\mathcal{E}\left(  \Psi\right)  \text{ for some
}\sigma\in\mathbb{R} \label{mm}%
\end{equation}
where%
\[
\mathfrak{M}_{\sigma}=\left\{  \Psi\in X:\mathcal{H}(\Psi)=\sigma\right\}
\]
Moreover $\Psi_{0}$ is called hylomorphic soliton if it satisfies definition
\ref{ds} namely if
\begin{equation}
\Gamma_{\sigma}=\left\{  \Psi_{0}\in X:\mathcal{E}\left(  \Psi\right)
=c_{\sigma}\right\}  \label{gs}%
\end{equation}
is a finite dimensional stable manifold.
\end{definition}

Observe that the energy and the charge are constant of the motion, then
$\Gamma_{\sigma}$ is an invariant set. Then the above definition is consistent
with definition \ref{ds}.

Let $\Psi_{0}$ be a hylomorphic stationary wave as in Def. \ref{hys} and set%
\begin{equation}
V(\Psi)=(\mathcal{E}\left(  \Psi\right)  -c_{\sigma})^{2}+(\mathcal{H}\left(
\Psi\right)  -{\sigma})^{2}; \label{liafun}%
\end{equation}
then we have that

\begin{prop}
Let $\Psi_{0}\ $and $\Gamma_{\sigma}$ be as in Def. \ref{hys} and let
$\Psi_{n}$ be any sequence in the phase space $X\ $which is supposed a metric
space with metric $d.$ Then if%
\begin{equation}
V(\Psi_{n})\rightarrow0\Rightarrow d(\Psi_{n},\Gamma_{\sigma})\rightarrow0
\label{alex}%
\end{equation}
$\Psi_{0}$ is a hylomorphic soliton.
\end{prop}

\textbf{Proof.} By (\ref{alex}), it is immediate to check that $V$ is a
Liapunov function relative to the flow $U_{t}$ defined by \ref{flusso}. Then,
by the well known Liapunov theorem, it follows that $\Gamma_{\sigma}$ is
stable and by the definitions \ref{dos} and \ref{ds} the conclusion follows.

$\square$

\bigskip

If $\Psi_{0}$ is a stationary wave of a dynamical system relative to a
Lagrangian which is invariant for the Lorentz or the Galileo group, then it is
possible to obtain a travelling wave just making a Lorentz boost (see
(\ref{LT})) or a Galilean transformation (see (\ref{galgr})). More precisely,
let $T_{\mathbf{v}}$ be the representation of a Lorentz boost (or a Galilean
transformation) relative to our system and let
\[
\Psi(t,x)=U_{t}\Psi_{0}(x)
\]
be the evolution of our stationary wave $\Psi_{0}(x);$ then%
\[
\Psi^{\prime}(t^{\prime},x^{\prime})=T_{\mathbf{v}}\Psi(t,x)
\]
is a solution of our equation which moves in time; then $\Psi_{\mathbf{v}%
}(x):=\Psi^{\prime}(t^{\prime},x^{\prime})|_{t=0}$ is a travelling wave at the
time $t=0.$ At pag. \pageref{sgsh} and pag. \pageref{solitone}, we will see as
this principle works in some particular cases.

Obviously, if $\Psi_{0}(x)$ is a hylomorphic standing wave $\Psi_{\mathbf{v}%
}(x)$ will be called hylomorphic travelling wave; moreover, if $\Psi_{0}(x) $
is orbitally stable, also $\Psi_{\mathbf{v}}(x)$ is orbitally stable and hence
is a hylomorphic soliton.

\subsection{Structure of hylomorphic solitons}

Now, as it happens in Th. \ref{noether}, we assume $\mathcal{E}$ and
$\mathcal{H}$ to be local quantities, namely, given $\Psi\in X,$ there exist
the density functions $\rho_{\mathcal{E},\Psi}\left(  x\right)  $ and
$\rho_{\mathcal{H},\Psi}\left(  x\right)  \in L^{1}(\mathbb{R}^{N})$ such that
\
\begin{align*}
\mathcal{E}\left(  \Psi\right)   &  =\int\rho_{\mathcal{E},\Psi}\left(
x\right)  \ dx\\
\mathcal{H}\left(  \Psi\right)   &  =\int\rho_{\mathcal{H},\Psi}\left(
x\right)  \ dx
\end{align*}

Energy and hylenic density allow to define the density of \textit{binding
energy }as follows:
\begin{equation}
\beta(t,x)=\beta_{\Psi}(t,x)=\left[  E_{0}\cdot\left\vert \rho_{\mathcal{H}%
,\Psi}\left(  t,x\right)  \right\vert -\rho_{\mathcal{E},\Psi}\left(
t,x\right)  \right]  ^{+} \label{ben}%
\end{equation}

The support of the binding energy density is called \textit{bound matter
region; }more precisely we have the following definition

\begin{definition}
\label{supsol}Given any configuration $\Psi$, we define the \textbf{bound
matter region} as follows
\[
\Sigma\left(  \Psi\right)  =\overline{\left\{  x:\beta_{\Psi}(t,x)\neq
0\right\}  }.
\]
If $\Psi_{0}$ is a soliton, the set $\Sigma\left(  \Psi_{0}\right)  $ is
called \textbf{support of the soliton}\emph{\ }at time\emph{\ }$t$.
\end{definition}

Thus a hylomorphic soliton $\Psi_{0}$ consists of bound matter localized in a
precise region of the space, namely $\Sigma\left(  \Psi_{0}\right)  $. This
fact gives the name to this type of soliton from the Greek words
\textquotedblright\textit{hyle}\textquotedblright=\textquotedblright%
\textit{matter}\textquotedblright\ and \textquotedblright\textit{morphe}%
\textquotedblright=\textquotedblright\textit{form}\textquotedblright.

\bigskip

We now set%

\begin{equation}
E_{0}=\;\underset{\varepsilon\rightarrow0}{\lim}\;\underset{\Psi\in
X_{\varepsilon}}{\inf}\frac{\mathcal{E}\left(  \Psi\right)  }{\left\vert
\mathcal{H}\left(  \Psi\right)  \right\vert } \label{brutta}%
\end{equation}
where
\begin{equation}
X_{\varepsilon}=\left\{  \Psi\in X:\forall x,\ \left\vert \Psi(x)\right\vert
_{V}<\varepsilon\right\}  . \label{icsep}%
\end{equation}

Now suppose that there is a state $\Psi$ which satisfies the inequality%
\begin{equation}
\frac{\mathcal{E}\left(  \Psi\right)  }{\left\vert \mathcal{H}\left(
\Psi\right)  \right\vert }<E_{0} \label{yc}%
\end{equation}
which will be called \textit{hylomorphy condition}. The quantity
\begin{equation}
\Lambda\left(  \Psi\right)  =\frac{\mathcal{E}\left(  \Psi\right)
}{\left\vert \mathcal{H}\left(  \Psi\right)  \right\vert }, \label{lambda}%
\end{equation}

\noindent is also an invariant of motion and will be called \textit{hylomorphy
ratio.}

Notice that, by def. (\ref{brutta}) and (\ref{lambda}), for any hylomorphic
solitary wave $\Psi_{0},$ we have that
\[
\Lambda\left(  \Psi_{0}\right)  =\frac{c_{\sigma}}{\sigma}\leq\frac{E_{0}%
}{\sigma}%
\]
However, if (\ref{yc}) holds, we have that
\[
\Lambda\left(  \Psi_{0}\right)  <\frac{E_{0}}{\sigma}%
\]
Actually the hylomorphy condition (\ref{yc}) seems to be a necessary condition
in order to have hylomorphic solitons.

The hylomorphy condition (\ref{yc}) guarantees the presence of bound matter
even if no soliton is present:

\begin{proposition}
\label{laura}If $\ \Lambda\left(  \Psi(0,\cdot)\right)  <E_{0}$, then for all
$t\in\mathbb{R}$
\[
\Sigma(\Psi(t,\cdot))\neq\varnothing
\]

\end{proposition}

\bigskip

\textbf{Proof.} To fix the ideas, assume $\mathcal{H}\left(  \Psi\right)
>0$.
\begin{align*}
\int\beta(t,x)  &  =\int\left[  \left\vert E_{0}\rho_{\mathcal{H,}U_{t}\Psi
}\left(  x\right)  \right\vert -\rho_{\mathcal{E,}U_{t}\Psi}\left(  x\right)
\right]  ^{+}\\
&  \geq\int E_{0}\rho_{\mathcal{H,}U_{t}\Psi}\left(  x\right)  -\rho
_{\mathcal{E,}U_{t}\Psi}\left(  x\right) \\
&  =E_{0}\mathcal{H}\left(  U_{t}\Psi\right)  -\mathcal{E}\left(  U_{t}%
\Psi\right) \\
&  =E_{0}\mathcal{H}\left(  \Psi\right)  -\mathcal{E}\left(  \Psi\right) \\
&  =\mathcal{H}\left(  \Psi\right)  \left[  E_{0}-\frac{\mathcal{E}\left(
\Psi\right)  }{\mathcal{H}\left(  \Psi\right)  }\right] \\
&  =\mathcal{H}\left(  \Psi\right)  \left[  E_{0}-\Lambda\left(  \Psi\right)
\right]  >0
\end{align*}

$\square$

\ 

If $\Psi(x)$ is a finite energy field usually it disperses as time goes on,
namely
\[
\underset{t\rightarrow\infty}{\lim}\left\Vert U_{t}\Psi(x)\right\Vert
_{L^{\infty}(\mathbb{R}^{N},V)}=0
\]
However, if $\Lambda\left(  \Psi\right)  <E_{0},$ this is not the case:

\bigskip

\begin{proposition}
If $\Lambda\left(  \Psi\right)  <E_{0},\;$then
\[
\underset{t\rightarrow\infty}{\min\lim}\left\Vert U_{t}\Psi\right\Vert
_{L^{\infty}(\mathbb{R}^{N},V)}=\delta>0
\]

\end{proposition}

\textbf{Proof}: To fix the idea, set $\Lambda\left(  \Psi\right)
=E_{0}-a,\;a>0.$ We argue indirectly and assume that, for every $\varepsilon
>0,$ there exists $\bar{t}$ such that
\[
\left\Vert U_{\bar{t}}\Psi\right\Vert _{L^{\infty}(\mathbb{R}^{N}%
,V)}<\varepsilon
\]
namely, $U_{\bar{t}}\Psi\in X_{\varepsilon}$ where $X_{\varepsilon}$ is
defined by (\ref{icsep}). Then, by (\ref{brutta}), if $\varepsilon$ is
sufficiently small
\[
\Lambda\left(  U_{\bar{t}}\Psi\right)  =\frac{\mathcal{E}\left(  U_{\bar{t}%
}\Psi\right)  }{\left\vert \mathcal{H}\left(  U_{\bar{t}}\Psi\right)
\right\vert }\geq E_{0}-\frac{a}{2}%
\]
Since $\Lambda\left(  U_{\bar{t}}\Psi\right)  =\Lambda\left(  \Psi\right)  $
we get a contradiction.

$\square$

\bigskip

Thus if $\Lambda\left(  \Psi\right)  <E_{0},$ by the above propositions the
field $\Psi$ and the bond matter field $\beta_{\Psi}$ will not disperse but
will form bumps of matter which eventually might lead to the formation of one
or more hylomorphic soliton. This remark shows the importance of the
hylomorphy condition (\ref{yc}).

\bigskip

\subsection{The \textit{swarm interpretation} of hylomprphic
solitons\label{inter}}

\bigskip

Clearly the physical interpretation of hylomorphic solitons depends on the
model which we are considering. However we can always assume a
\textit{conventional interpretations }which we will call \textit{swarm
interpretation }since the soliton is regarded as a \textit{swarm }of particles
bound together\textit{. }In each particular physical situation this
interpretation, might have or might not have any physical meaning; in any case
it represents a pictorial way of thinking of the mathematical phenomena which
occur. This interpretation is consistent with the names and the definitions
given in the previous section.

\bigskip

We assume that $\Psi$ is a field which describes a fluid consisting of
particles; the particles density is given by the function $\rho_{\mathcal{H}%
}(t,x)=\rho_{\mathcal{H},\Psi}(t,x)$ which, of course satisfies a continuity
equation
\begin{equation}
\partial_{t}\rho_{\mathcal{H}}+\nabla\cdot\mathbf{J}_{\mathcal{H}}=0
\label{CE}%
\end{equation}
where $\mathbf{J}_{\mathcal{H}}$ is the flow of particles. Hence $\mathcal{H}
$ is the total number of particles. Notice that $\mathcal{H}$ does not need to
be an integer number and you may assume that fractional particle can exist or
that $\mathcal{H}$ is a sort of limit valid for a very large number of
particle as it happens in fluid-dynamics. Also, in some equations as for
example in NKG, $\mathcal{H}$ can be negative; in this case we assume the
existence of antiparticle.

Thus, the hylomorphy ratio
\[
\Lambda\left(  \Psi\right)  =\frac{\mathcal{E}\left(  \Psi\right)
}{\left\vert \mathcal{H}\left(  \Psi\right)  \right\vert }%
\]
represents the average energy of each particle (or antiparticle). The number
$E_{0}$ defined by (\ref{brutta}) is interpreted as the rest energy of each
particle when they do not interact with each other. If $\Lambda\left(
\Psi\right)  >E_{0},$ then the average energy of each particle is bigger than
the rest energy; if $\Lambda\left(  \Psi\right)  <E_{0},$ the opposite occurs
and this fact means that particles act with each other with an attractive force.

If the particles would be at rest and they would not act on each other, their
energy density would be
\[
E_{0}\cdot\left\vert \rho_{\mathcal{H}}(t,x)\right\vert ;
\]
but their energy density is $\rho_{\mathcal{E}}(t,x);$ if
\[
\rho_{\mathcal{E}}(t,x)<E_{0}\cdot\left\vert \rho_{\mathcal{H}}%
(t,x)\right\vert ;
\]
then, in the point $x$ at time $t,$ the particles attract each other with a
force which is stronger that the repulsive forces; this explains the name
\textit{density of} \textit{bond energy }given to $\beta(t,x)$ in (\ref{ben}).

Thus a soliton relative to the state $\Psi$ can be considered as a "rigid"
object occupying the region of space $\Sigma\left(  \Psi\right)  $ (cf. Def.
\ref{supsol}); it consists of particles which stick with each other; the
energy to destroy the soliton is given by
\[
\int\beta_{\Psi}(t,x)dx=\int_{\Sigma\left(  \Psi\right)  }\left(
E_{0}\left\vert \rho_{\mathcal{H}}(t,x)\right\vert -\rho_{\mathcal{E}%
}(t,x)\right)  dx
\]

However out of $\Sigma\left(  \Psi\right)  $ the energy density is bigger than
$E_{0}\cdot\left\vert \rho_{\mathcal{H}}(t,x)\right\vert ;$ thus the total
energy necessary to reduce the soliton to isolated particles is give by
\[
E_{0}\cdot\left\vert \mathcal{H}\left(  \Psi\right)  \right\vert
-\mathcal{E}\left(  \Psi\right)  .
\]

As it is shown in Prop. \ref{laura}, there are states $\Psi$ such that
$\Sigma\left(  \Psi\right)  \neq0$ but they are not necessarily solitons. In
these states, $\Sigma\left(  \Psi\right)  $ is a region where the particles
stick with each other but they do not have reached a stable configurations;
the shape of $\Sigma\left(  \Psi\right)  $ might changes with time. In may
concrete situations, such states may evolve toward one or more solitons. In
these cases we say that the solitons are asymptotically stable. The study of
asymptotical stability is a problem quite involved. We refer to \cite{kom96},
\cite{bus03}, \cite{cucc08} and \cite{cucc08NS} and their references.

\section{The nonlinear Schroedinger equation\label{SSE}}

\subsection{General features of NS}

The Schroedinger equation for a particle which moves in a potential $V(x)$ is
given by
\[
i\frac{\partial\psi}{\partial t}=-\frac{1}{2}\Delta\psi+V(x)\psi
\]

We are interested to the nonlinear Schroedinger equation:
\begin{equation}
i\frac{\partial\psi}{\partial t}=-\frac{1}{2}\Delta\psi+\frac{1}{2}W^{\prime
}(\psi)+V(x)\psi\label{NSV}%
\end{equation}
where
\begin{equation}
W^{\prime}(\psi)=\frac{\partial W}{\partial\psi_{1}}+i\frac{\partial
W}{\partial\psi_{2}} \label{w'}%
\end{equation}
namely
\[
W^{\prime}(\psi)=F^{\prime}(\left\vert \psi\right\vert )\frac{\psi}{\left\vert
\psi\right\vert }.
\]
for some smooth function $F:\left[  0,\infty\right)  \rightarrow\mathbb{R}.$

We always assume that
\[
W(0)=W^{\prime}(0)=0
\]

If $V(x)=0,$ then we get the equation
\begin{equation}
i\frac{\partial\psi}{\partial t}=-\frac{1}{2}\Delta\psi+\frac{1}{2}W^{\prime
}(\psi); \tag{NS}\label{NS}%
\end{equation}
this equation can be considered as the simplest equation which is variational
and invariant for a representation of the Galileo group.

First of all let us check that it is variational:

\begin{proposition}
Equation (\ref{NSV}) is the Euler-Lagrange equation relative to the Lagrangian
density
\begin{equation}
\mathcal{L}=\operatorname{Re}\left(  i\partial_{t}\psi\overline{\psi}\right)
-\frac{1}{2}\left\vert \nabla\psi\right\vert ^{2}-W\left(  \psi\right)
-V(x)\left\vert \psi\right\vert ^{2} \label{lagr}%
\end{equation}

\end{proposition}

\textbf{Proof} Set
\begin{align*}
\mathcal{S}(\psi)  &  =\mathcal{S}_{1}(\psi)+\mathcal{S}_{2}(\psi)\\
\mathcal{S}_{1}(\psi)  &  =\int\operatorname{Re}\left(  i\partial_{t}%
\psi\overline{\psi}\right)  dxdt;\ \mathcal{S}_{2}(\psi)=-\int\left[  \frac
{1}{2}\left\vert \nabla\psi\right\vert ^{2}+W\left(  \psi\right)
+V(x)\left\vert \psi\right\vert ^{2}\right]  dxdt
\end{align*}
and set $\psi=u_{1}+iu_{2}.$ We have%
\begin{align*}
\mathcal{S}_{1}(\psi)  &  =\int\operatorname{Re}\left(  i\partial_{t}%
\psi\overline{\psi}\right)  dxdt\\
&  =\int\operatorname{Re}\left[  \left(  i\partial_{t}u_{1}-\partial_{t}%
u_{2}\right)  \left(  u_{1}-iu_{2}\right)  \right]  dxdt\\
&  =\int\left(  \partial_{t}u_{1}u_{2}-\partial_{t}u_{2}u_{1}\right)  dxdt
\end{align*}
Then, if $\varphi=v_{1}+iv_{2}$
\begin{align*}
d\mathcal{S}_{1}(\psi)\left[  \varphi\right]   &  =\int\left(  \partial
_{t}u_{1}v_{2}+\partial_{t}v_{1}u_{2}-\partial_{t}u_{2}v_{1}-\partial_{t}%
v_{2}u_{1}\right)  dxdt\\
&  =\int\left(  2\partial_{t}u_{1}v_{2}-2\partial_{t}u_{2}v_{1}\right)  dxdt\\
&  =\int\operatorname{Re}\left[  2i\left(  \partial_{t}u_{1}+\partial_{t}%
u_{2}\right)  \left(  v_{1}-iv_{2}\right)  \right] \\
&  =\int\operatorname{Re}\left(  2i\partial_{t}\psi\overline{\varphi}\right)
\end{align*}

Then
\begin{align*}
d\mathcal{S}_{2}(\psi)\left[  \varphi\right]   &  =-\int\left[
\operatorname{Re}\left\langle \nabla\psi,\nabla\varphi\right\rangle
+\operatorname{Re}\left(  W^{\prime}\left(  \psi\right)  \overline{\varphi
}\right)  +2\operatorname{Re}\left(  V(x)\psi\overline{\varphi}\right)
\right]  dxdt\\
&  =-\int\left[  \operatorname{Re}\left(  -\Delta\psi\overline{\varphi
}\right)  +\operatorname{Re}\left(  W^{\prime}\left(  \psi\right)
\overline{\varphi}\right)  +2\operatorname{Re}\left(  V(x)\psi\overline
{\varphi}\right)  \right]  dxdt\\
&  =-\int\operatorname{Re}\left[  \left(  -\Delta\psi+W^{\prime}\left(
\psi\right)  +2V(x)\psi\right)  \overline{\varphi}\right]  dxdt.
\end{align*}%
\[
d\mathcal{S}(\psi)\left[  \varphi\right]  =\int\operatorname{Re}\left[
\left(  2i\partial_{t}\psi+\Delta\psi+W^{\prime}\left(  \psi\right)
+2V(x)\psi\right)  \overline{\varphi}\right]  dxdt
\]
So, the critial points of $\mathcal{S},$ satisy the equation
\[
2i\partial_{t}\psi+\Delta\psi-W^{\prime}\left(  \psi\right)  -2V(x)\psi=0
\]
which is equivalent to (\ref{NSV}).

$\square$

Sometimes it is useful to write $\psi$ in polar form
\begin{equation}
\psi(t,x)=u(t,x)e^{iS(t,x)}. \label{polar1}%
\end{equation}
where $u(t,x)\in\mathbb{R}^{+}$ and $S(t,x)\in\mathbb{R}/\left(
2\pi\mathbb{Z}\right)  $. Thus the state of the system ${\psi}$ is uniquely
defined by the couple of variables $(u,S)$. Using these variables, the action
$\mathcal{S=}\int\mathcal{L}dxdt$ takes the form
\[
\mathcal{S}(u,S)=-\int\left[  \frac{1}{2}\left\vert \nabla u\right\vert
^{2}+W(u)+\left(  \partial_{t}S+\frac{1}{2}\left\vert \nabla S\right\vert
^{2}+V(x)\right)  u^{2}\right]  dx
\]
and equation (\ref{NSV}) becomes:%

\begin{equation}
-\frac{1}{2}\Delta u+\frac{1}{2}W^{\prime}(u)+\left(  \partial_{t}S+\frac
{1}{2}\left\vert \nabla S\right\vert ^{2}+V(x)\right)  u=0 \label{Sh1}%
\end{equation}

\begin{equation}
\partial_{t}\left(  u^{2}\right)  +\nabla\cdot\left(  u^{2}\nabla S\right)  =0
\label{Sh2}%
\end{equation}

\subsection{First integrals of NS and the hylenic ratio}

The results of section \ref{cl} and easy computations show that the integral
of motion of \ref{NSV} are given by the following expression:

\begin{itemize}
\item \textbf{Energy}. We have
\begin{equation}
\mathcal{E}(\psi)=\int\left[  \frac{1}{2}\left\vert \nabla\psi\right\vert
^{2}+W(\psi)+V(x)\left\vert \psi\right\vert ^{2}\right]  dx
\end{equation}

\end{itemize}

Using (\ref{polar1}) we get:
\begin{equation}
\mathcal{E}(\psi)=\int\left(  \frac{1}{2}\left\vert \nabla u\right\vert
^{2}+W(u)\right)  dx+\int\left(  \frac{1}{2}\left\vert \nabla S\right\vert
^{2}+V(x)\right)  u^{2}dx \label{Schenergy}%
\end{equation}

\begin{itemize}
\item \textbf{Momentum. }The momentum is constant in time if the Lagrangian is
space-traslation invariant; this happens when $V$ is a constant. In
particular, in eq. \ref{NS}, we have that
\begin{equation}
\mathbf{P}=\operatorname{Im}\int\nabla\psi\overline{\psi}\;dx \label{Psch}%
\end{equation}

Using (\ref{polar1}) we get:
\begin{equation}
\mathbf{P}=\int u^{2}\nabla S\;dx
\end{equation}

\item \textbf{Angular momentum. }If we assume that $V$ is a constant, the
angular momentum is an integral of motion and, in \ref{NS}, we have
\begin{equation}
\mathbf{M}=\operatorname{Im}\int\mathbf{x}\times\nabla\psi\overline{\psi}\;dx
\end{equation}
Using (\ref{polar1}) we get:
\begin{equation}
\mathbf{M}=\int\mathbf{x}\times\nabla S\,u^{2}\;dx
\end{equation}

\item \textbf{Hylenic charge.} Here the hylenic charge has the following
expression
\begin{equation}
\mathcal{H}(\psi)=\int\left\vert \psi\right\vert ^{2}dx=\int u^{2}dx
\label{Hsch}%
\end{equation}

\item \textbf{Barycenter (or Hylecenter) velocity. }The quantity preserved by
the Galileo transformation (\ref{gal}) is the following
\begin{equation}
\mathbf{K}=\int\mathbf{x}u^{2}dx-t\mathbf{P} \label{leann}%
\end{equation}
Here we make the explicit computation since it is more difficult than the
computations for the other first integrals.
\end{itemize}

\textbf{Proof. }Let us compute $K_{1}$ using Th. \ref{noe}; in this case the
parameter $\lambda$ is $v$ the first component of the velocity $\mathbf{v=(}%
v,0,0\mathbf{)}$ which appears in (\ref{galgr}). In this case the computations
are easier if we use polar coordinates: we have
\[
\mathcal{L}=-\left[  \frac{1}{2}\left\vert \nabla u\right\vert ^{2}%
+W(u)+\left(  \partial_{t}S+\frac{1}{2}\left\vert \nabla S\right\vert
^{2}\right)  u^{2}\right]
\]
Moreover we have that%
\begin{align*}
\mathbf{x}_{v}  &  =\mathbf{x}-\mathbf{v}t\\
t_{v}  &  =t
\end{align*}
and recalling (\ref{gal}), we have that the representation $T_{g_{v}}$ acts on
a state $\psi$ of the system as follows%
\[
\left(  T_{g_{v}}\psi\right)  (t,x)=\psi\left(  t,x-\mathbf{v}t\right)
e^{i(\mathbf{v\cdot}x-\frac{1}{2}v^{2}t)}=u\left(  t,x-\mathbf{v}t\right)
e^{iS(t,x-\mathbf{v}t)}e^{i(\mathbf{v\cdot}x-\frac{1}{2}v^{2}t)}%
\]
namely%
\begin{align*}
u_{v}(t,x)  &  =u\left(  t,x-\mathbf{v}t\right) \\
S_{v}(t,x)  &  =S\left(  t,x-\mathbf{v}t\right)  +vx_{1}-\frac{1}{2}v^{2}t
\end{align*}

Then, by (\ref{bellissima})%
\begin{align*}
\rho_{K_{1}}  &  =\left[  \frac{\partial\mathcal{L}}{\partial u_{t}}%
\frac{\partial u_{v}}{\partial v}+\frac{\partial\mathcal{L}}{\partial S_{t}%
}\frac{\partial S_{v}}{\partial v}-\mathcal{L}\frac{\partial t_{v}}{\partial
v}\right]  _{v=0}=\left[  u_{v}^{2}\cdot\frac{\partial S_{v}}{\partial
v}\right]  _{v=0}\\
&  =\left[  u_{v}^{2}\left(  -\frac{\partial S_{v}}{\partial x_{1}}%
\ t+x_{1}-vt\right)  \right]  _{v=0}=\left(  x_{1}-t\frac{\partial S_{v}%
}{\partial x_{1}}\right)  u^{2}%
\end{align*}

Thus%
\[
K_{1}=\int x_{1}u^{2}dx-t\int\frac{\partial S_{v}}{\partial x_{1}}u^{2}dx=\int
x_{1}u^{2}dx-tP_{1}%
\]

$\square$

The three components of $\mathbf{K}$ are the integrals of motion relative to
the Galileo invariance. Let us interprete this fact in a more meaningful way.
If we derive both sides of (\ref{leann}) with respect to $t,$ we get
\begin{equation}
0=\frac{d}{dt}\left(  \int\mathbf{x}u^{2}dx\right)  -\mathbf{P} \label{leann2}%
\end{equation}

If we define the \textit{barycenter} (or \textit{hylecenter)} as follows
\begin{equation}
\mathbf{q:=}\frac{\int\mathbf{x}u^{2}dx}{\int u^{2}dx}=\frac{\int
\mathbf{x}u^{2}dx}{\mathcal{H}} \label{bary}%
\end{equation}

By (\ref{leann2}) we get%
\begin{equation}
\mathbf{\dot{q}}=\frac{\mathbf{P}}{\mathcal{H}}=const. \label{leann3}%
\end{equation}

Then, the three integrals of motion relative to the Galileo invariance imply
that the three components of $\mathbf{\dot{q}}$ are constant.

From now on, we will study the existence of solitary waves assuming $V(x)=0.$
If $W\in C^{2},$ we can write
\begin{equation}
W(s)=\frac{1}{2}\ as^{2}+N(s) \label{NO}%
\end{equation}
where $N^{\prime\prime}(s)=0.$

Now let us compute the limit (\ref{brutta})$.$

\begin{theorem}
\label{ttt}Assume that $W$ is given by (\ref{NO}) and that $E_{0}$ is given by
(\ref{brutta}). Then
\begin{equation}
E_{0}=\frac{1}{2}\ a. \label{lula}%
\end{equation}

\end{theorem}

\textbf{Proof}. We have that
\begin{align*}
\Lambda\left(  \Psi\right)   &  =\frac{\mathcal{E}(\Psi)}{\mathcal{H}\left(
\Psi\right)  }=\frac{\int\frac{1}{2}\left(  \left\vert \nabla u\right\vert
^{2}+\left\vert \nabla S\right\vert ^{2}u^{2}\right)  +W(\psi)dx}{\int
u^{2}dx}\\
&  \geq\frac{\int\frac{1}{2}au^{2}+N(u)dx}{\int u^{2}dx}=\frac{1}{2}%
a+\frac{\int N(u)dx}{\int u^{2}dx}%
\end{align*}
Then since $N(u)=O(u^{3})$ for $u\rightarrow0,$ we have that
\[
\Lambda\left(  \Psi\right)  =\ \underset{\varepsilon\rightarrow0}{\lim
}\;\underset{\Psi\in X_{\varepsilon}}{\inf}\frac{\mathcal{E}\left(
\Psi\right)  }{\mathcal{H}\left(  \Psi\right)  }\geq\frac{1}{2}a.
\]
In order to prove the opposite inequality, take $\Psi_{\varepsilon
,R}=\varepsilon u_{R}$\ where $u_{R}(x)$ is defined as follows
\[
u_{R}(x)=\left\{
\begin{array}
[c]{cc}%
1 & if\;\;|x|<R\\
0 & if\;\;|x|>R+1\\
1+R-|x| & if\;\;R<|x|<R+1
\end{array}
\right.
\]

Then
\begin{align*}
\underset{\Psi\in X_{\varepsilon}}{\inf}\frac{\mathcal{E}\left(  \Psi\right)
}{\mathcal{H}\left(  \Psi\right)  }  &  \leq\frac{\mathcal{E}\left(
\Psi_{\varepsilon,R}\right)  }{\mathcal{H}\left(  \Psi_{\varepsilon,R}\right)
}=\frac{\varepsilon^{2}\int\left[  \frac{1}{2}\left\vert \nabla u_{R}%
\right\vert ^{2}+\frac{1}{2\varepsilon^{2}}W(\varepsilon u)\right]
dx}{\varepsilon^{2}\int\,u_{R}^{2}dx}\\
&  =\frac{\int\left[  \frac{1}{2}\left\vert \nabla u_{R}\right\vert ^{2}%
+\frac{1}{2}au_{R}^{2}+\frac{1}{2\varepsilon^{2}}N(\varepsilon u)\right]
dx}{\int\,u_{R}^{2}dx}\\
&  \leq\frac{1}{2}a+\frac{1}{2}\frac{\int\left\vert \nabla u_{R}\right\vert
^{2}dx}{\int\,u_{R}^{2}dx}+\frac{\int\frac{1}{2\varepsilon^{2}}N(\varepsilon
u)dx}{\int\,u_{R}^{2}dx}\\
&  =1+O\left(  \frac{1}{R}\right)  +O\left(  \varepsilon\right)
\end{align*}
and so, we have that
\[
\underset{\Psi\in X_{\varepsilon}}{\inf}\frac{\mathcal{E}\left(  \Psi\right)
}{\mathcal{H}\left(  \Psi\right)  }\leq\frac{1}{2}a
\]

$\square$

By (\ref{NO}) and (\ref{lula}), we can write
\begin{equation}
W(s)=E_{0}s^{2}+N(s),\ \ N(0)=N^{\prime}(0)=N^{\prime\prime}(0)=0
\label{fodiw}%
\end{equation}

It is not restrictive to assume that
\[
E_{0}=0
\]
In fact, if $\psi_{0}(t,x)$ is a solution of the equation
\begin{equation}
i\frac{\partial\psi}{\partial t}=-\frac{1}{2}\Delta\psi+E_{0}\psi+\frac{1}%
{2}N^{\prime}(\psi) \label{ale}%
\end{equation}
then
\[
\psi_{1}(t,x)=\psi_{0}(t,x)e^{-iE_{0}t}%
\]
satisfies the equation
\begin{equation}
i\frac{\partial\psi}{\partial t}=-\frac{1}{2}\Delta\psi+\frac{1}{2}N^{\prime
}(\psi) \label{ase}%
\end{equation}

So equation (\ref{ale}) and (\ref{ase}) have the same dynamics except of a
phase factor. This fact, in the swarm model, can be interpreted saying that
the internal energy of a particle (namely $E_{0}$) does not affect the
dynamics of the particle.

\bigskip

\subsection{Swarm interpretation of NS\label{sins}}

Before giving the swarm interpretation to equation \ref{NS}, we will write it
with the usual physical constants $m$ and $\hslash:$%
\begin{equation}
i\hslash\frac{\partial\psi}{\partial t}=-\frac{\hslash^{2}}{2m}\Delta
\psi+\frac{1}{2}W^{\prime}(\psi)+V(x)\psi\label{NSCON}%
\end{equation}
Here $m$ has the dimension of \textit{mass} and $\hslash,$ the Plank constant,
has the dimension of \textit{action}.

The polar form of a state $\psi$ is written as follows
\begin{equation}
\psi(t,x)=u(t,x)e^{iS(t,x)/\hslash} \label{polar2}%
\end{equation}
and equations (\ref{Sh1}) and (\ref{Sh2}) become%

\begin{equation}
-\frac{\hslash^{2}}{2m}\Delta u+\frac{1}{2}W^{\prime}(u)+\left(  \partial
_{t}S+\frac{1}{2m}\left\vert \nabla S\right\vert ^{2}+V(x)\right)  u=0
\label{Sh1c}%
\end{equation}

\begin{equation}
\partial_{t}\left(  u^{2}\right)  +\nabla\cdot\left(  u^{2}\frac{\nabla S}%
{m}\right)  =0 \label{Sh2c}%
\end{equation}

The continuity equation (\ref{CE}) for \ref{NS} is given by (\ref{Sh2c}). This
equation allows us to interprete the matter field to be a fluid composed by
particles whose density is given by
\[
\rho_{\mathcal{H}}=u^{2}%
\]
and which move in the velocity field
\begin{equation}
\mathbf{v}=\frac{\nabla S}{m}. \label{vel0}%
\end{equation}
So equation (\ref{Sh2c}) reads
\[
\partial_{t}\rho_{\mathcal{H}}+\nabla\cdot\left(  \rho_{\mathcal{H}}%
\mathbf{v}\right)  =0
\]

If
\begin{equation}
-\frac{\hslash^{2}}{2m}\Delta u+\frac{1}{2}W^{\prime}(u)\ll u, \label{rosaS}%
\end{equation}
equation (\ref{Sh1c}) can be approximated by
\begin{equation}
\partial_{t}S+\frac{1}{2m}\left\vert \nabla S\right\vert ^{2}+V(x)=0.
\label{hjS}%
\end{equation}
This is the Hamilton-Jacobi equation of a particle of mass $m$ in a potential
field $V$ (cf. eq. \ref{hjN}). The trajectory $q(t)$ of each particle, by
(\ref{vel0}) satisfies the equation
\begin{equation}
\dot{q}=\frac{\nabla S}{m} \label{velq0}%
\end{equation}
(cf. eq. \ref{hjN2}).

If we do not assume (\ref{rosaS}), equation (\ref{hjS}) needs to be replaced
by
\begin{equation}
\partial_{t}S+\frac{1}{2m}\left\vert \nabla S\right\vert ^{2}+V+Q(u)=0
\label{hjqS}%
\end{equation}
with
\begin{equation}
Q(u)=\frac{-\left(  \hslash^{2}/m\right)  \Delta u+W^{\prime}(u)}{2u}%
=E_{0}+\frac{-\left(  \hslash^{2}/m\right)  \Delta u+N^{\prime}(u)}{2u}
\label{Qu}%
\end{equation}

The term $Q(u)$ can be regarded as a field describing a sort of interaction
between particles. The term $E_{0}$ does not affect the dynamics, while the
other terms might contribute to the formations of hylomorphic solitons.

\ Given a wave of the form (\ref{polar2}), the local frequency and the local
wave number are defined as follows:
\begin{align*}
\omega(t,x)  &  =-\frac{\partial_{t}S(t,x)}{\hslash}\\
\mathbf{k}(t,x)  &  =\frac{\nabla S(t,x)}{\hslash};
\end{align*}
the energy of each particle moving according to (\ref{hjS}), is given by
\[
E=-\partial_{t}S
\]
and its momentum is given by
\[
\mathbf{p}=\nabla S;
\]
thus we have that
\begin{align*}
E  &  =\hslash\omega\\
\mathbf{p}  &  =\hslash\mathbf{k};
\end{align*}
these two equations are the De Broglie relation. It is interesting to observe
that they have be deduced by the swarm interpretation of the Schroedinger equation.

\bigskip

We recall again that the swarm interpretation is just a useful pictorial way
to look at to NS. In physical models, in general, there are different
interpretations. Here we will mention very shortly some of them.

In the traditional model of quantum mechanics for one particle, we have $W=0$
and $\rho_{\mathcal{H}}$ is interpreted as a probability density of the
position of this particle.

One of the most important applications of NS is in the \textit{Bose-Einstein
condensate}. In this case $W(\psi)=\frac{U_{0}}{4}\left\vert \psi\right\vert
^{4}$ (where $U_{0}$ is a constant) and \ref{NS} takes the name of
\textit{Gross-Pitaevski equation}. Here $\left\vert \psi\right\vert ^{2}$ is
interpreted as the particle density which in this case are bosons (as, for
example, atoms).

\subsection{An existence result for an elliptic equation\label{giovanna}}

Many existence theorems of solitary waves reduce to the following elliptic
equation in $\mathbb{R}^{N}$
\begin{equation}
\left\{
\begin{array}
[c]{ll}%
-\Delta u+G^{\prime}(u)=0 & \\
\ u>0 &
\end{array}
\right.  \label{gianna}%
\end{equation}
where $G$ is a $C^{1}$-function with $G(0)=0.$

This equation has been studied by many authors (see e.g. \cite{strauss},
\cite{coleman78}, \cite{BL81} and their bibliography). In particular, in
\cite{BL81}, there are necessary and "almost sufficient" conditions for the
existence of "finite energy" solutions.

Since this equation is the basic equation for the existence of solitary waves,
we give an existence proof. This proof is a variant of the proof in
\cite{BL81}. It is simpler but it uses slightly more restrictive assumptions.

Equation (\ref{gianna}) is the Euler-Lagrange equation relative to the
functional
\[
J(u)=\frac{1}{2}\int\left\vert \nabla u\right\vert ^{2}dx+\int G(u)dx,
\]

We assume that $G:\mathbb{R}^{+}\rightarrow\mathbb{R}$ satisfies the following assumptions:

\begin{itemize}
\item (G-i)$\ \ G(0)=G^{\prime}(0)=0.$

\item (G-ii) \ $G^{\prime}(s)\geq c_{1}s-c_{2}s^{p-1},\ \ 2< p<2^{\ast}, $
$s>0,\ c_{1},c_{2}>0;$

\item (G-iii)$\ \ \exists s_{0}\in\mathbb{R}^{+}:\;G(s_{0})<0.$
\end{itemize}

\begin{theorem}
\label{pippo}Assume that $G$ satisfies (G-i), (G-ii), (G-iii). Then eq.
(\ref{gianna}) has a nontrivial finite energy solution.
\end{theorem}

In order to prove the above theorem, first of all we define an auxiliary
function $\bar{G}$ as follows:

If
\begin{equation}
\left\vert G^{\prime}(s)\right\vert \leq c_{3}+c_{4}\left\vert s\right\vert
^{p-1} \label{b33}%
\end{equation}
(where $p$ is defined by (G-ii)), we set
\[
\bar{G}(s)=\left\{
\begin{array}
[c]{lll}%
G\left(  s\right)  & for & s\geq0\\
0 & for & s\leq0
\end{array}
\right.
\]
If $G$ does not satisfy (\ref{b33}), then, by (G-ii), there exist $s_{1}%
>s_{0},$ such that $G^{\prime}(s_{1})>0.$ In this case we set
\[
\bar{G}(s)=\left\{
\begin{array}
[c]{lll}%
G\left(  s\right)  & for & 0\leq s\leq s_{1};\\
G\left(  s_{1}\right)  +G^{\prime}(s_{1})\left(  s-s_{1}\right)  & for & s\geq
s_{1};\\
0 & for & s\leq0.
\end{array}
\right.
\]

In any case, $\bar{G}$ satisfies the assumptions (G-i), (G-ii) (G-iii) and the
following ones:

\begin{itemize}
\item (G-iiii) \ $\left\vert \bar{G}^{\prime}(s)\right\vert \leq c_{3}%
+c_{4}\left\vert s\right\vert ^{p-1},\ \ 2\leq p<2^{\ast}$

\item (G-v) \ $\forall s<0,\ \bar{G}(s)=0$
\end{itemize}

\bigskip

\begin{lemma}
Let $u\in H^{1}$ be a solution of the following equation:
\begin{equation}
-\Delta u+\bar{G}^{\prime}(u)=0 \label{gia}%
\end{equation}
Then, $u$ is positive and it is a is a solution of (\ref{gianna}).
\end{lemma}

\textbf{Proof. }The fact that $u$ is positive is a straightforward consequence
of the maximum principle and the fact that $\bar{G}^{\prime}(u)=0$ for
$u\leq0.$ Then if (\ref{b33}) holds, $\bar{G}^{\prime}(u)=G^{\prime}(u)$ and
hence $u$ is a solution of (\ref{gianna}). If (\ref{b33}) does not hold, by
the maximum principle it follows that $u(x)\leq s_{1};$ then, also in this
case $\bar{G}^{\prime}(u)=G^{\prime}(u)$ and $u$ is a solution of
(\ref{gianna}).

$\square$

Thus, by the above lemma it is not restrictive to assume that $G$ satisfies
(G-i), (G-iii), (G-iiii) and (G-v), since otherwise we can work with $\bar{G}
$.

We set%
\[
H_{r}^{1}=\left\{  u\in H^{1}\left(  \mathbb{R}^{N}\right)  :u=u(|x|)\right\}
\]

\begin{lemma}
\label{ucap}There exists $\breve{u}\in H_{r}^{1}$ such that
\[
\int G(\breve{u})dx<0
\]

\end{lemma}

\textbf{Proof.} We set
\begin{equation}
u_{R}(x)=\left\{
\begin{array}
[c]{lll}%
s_{0} & for & \left\vert x\right\vert <R;\\
s_{0}-s_{0}(\left\vert x\right\vert -R) & for & R\leq\left\vert x\right\vert
\leq R+1;\\
0 & for & \left\vert x\right\vert >R+1.
\end{array}
\right.
\end{equation}
Thus we have
\begin{align*}
&  \int_{\mathbb{R}^{N}}G(u)dx\\
&  \leq\left[  \max_{|s|\in\lbrack R,R+1]}G(s)\right]  \int_{R}^{R+1}%
r^{N-1}dr+\int_{0}^{R}G(s_{0})r^{N-1}dr\\
&  \leq C_{1}\left[  \left(  R+1\right)  ^{N}-R^{N}\right]  +\frac{1}%
{N}G(s_{0})R^{N}\\
&  \leq C_{2}R^{N-1}+\frac{1}{N}G(s_{0})R^{N}%
\end{align*}
where $C_{1},C_{2}$ are positive constant. By (G-iii), .$G(s_{0})<0;$ hence,
for $R$ sufficiently large, $\int G(u_{R})dx<0$

$\square$

\begin{lemma}
\label{lstrauss}If $u\in\mathcal{D\ }$\footnote{Here $\mathcal{D}$ denotes the
space of $C^{\infty}$-functions with compact suport.}, is a radially symmetric
function, then%
\[
\left\vert u(x)\right\vert \leq C_{N}\frac{\left\Vert u\right\Vert _{H^{1}}%
}{\left\vert x\right\vert ^{\frac{N-1}{2}}}%
\]

\end{lemma}

\textbf{Proof.} We have that%

\[
\frac{d}{dr}\left(  r^{N-1}u^{2}\right)  =2r^{N-1}u\frac{du}{dr}+\left(
N-1\right)  r^{N-2}u^{2}\geq2r^{N-1}u\frac{du}{dr}%
\]
Then, integrating over $\left(  R,+\infty\right)  $ we get%
\[
-R^{N-1}u\left(  R\right)  ^{2}\geq2\int_{R}^{+\infty}r^{N-1}u\frac{du}{dr}dr
\]
and so%
\[
R^{N-1}u\left(  R\right)  ^{2}\leq2\int_{R}^{+\infty}\left\vert u\frac{du}%
{dr}\right\vert r^{N-1}dr\leq\int_{R}^{+\infty}\left(  \left\vert \frac
{du}{dr}\right\vert ^{2}+u^{2}\right)  r^{N-1}dr\leq C_{N} \left\Vert
u\right\Vert _{H^{1}}^{2}%
\]
$\square$

\bigskip

\begin{theorem}
\label{strauss}(Strauss \cite{strauss}) For $p\in\left(  2,2^{\ast}\right)  $
the embedding
\[
H_{r}^{1}\rightarrow L^{p}%
\]
is compact.
\end{theorem}

\textbf{Proof. }Let $u_{n}\rightharpoonup0$ weakly in $H_{r}^{1};$ we need to
prove that $u_{n}\longrightarrow0$ strong in $L^{p}.$ Since $u_{n}%
\rightharpoonup0,$ then there is a constant $M$ such that $\left\Vert
u_{n}\right\Vert _{H^{1}}\leq M;$ so by lemma \ref{lstrauss}, we have that
\begin{align}
\int_{\mathbb{R}^{N}-B_{R}}\left\vert u_{n}\right\vert ^{p}dx  &  \leq
||u_{n}||_{L^{\infty}\left(  \mathbb{R}^{N}-B_{R}\right)  }^{p-2}%
\int_{\mathbb{R}^{N}-B_{R}}\left\vert u_{n}\right\vert ^{2}dx\label{zita}\\
&  \leq\left(  \frac{\left\Vert u\right\Vert _{H^{1}}}{R^{\frac{N-1}{2}}%
}\right)  ^{p-2}\int_{\mathbb{R}^{N}-B_{R}}\left\vert u_{n}\right\vert
^{2}dx\\
&  \leq\left(  \frac{M}{R^{\frac{N-1}{2}}}\right)  ^{p-2}\left\Vert
u\right\Vert _{L^{2}}^{2}\leq\frac{M^{p}}{R^{\alpha}}%
\end{align}
where%
\[
\alpha=\frac{\left(  N-1\right)  \left(  p-2\right)  }{2}>0.
\]
Since $u_{n}\longrightarrow0$ strongly in $L^{p}\left(  B_{R}\right)  ,$ by
(\ref{zita}), we have that%
\[
\underset{n\rightarrow\infty}{\lim\inf}\int\left\vert u_{n}\right\vert
^{p}dx=~\underset{n\rightarrow\infty}{\lim\inf}||u_{n}||_{L^{p}\left(
B_{R}\right)  }^{p}+\underset{n\rightarrow\infty}{~\lim\inf}||u_{n}%
||_{L^{p}\left(  \mathbb{R}^{N}-B_{R}\right)  }^{p}\leq\frac{M^{p}}{R^{\alpha
}}%
\]
By the arbitrariness of $R,$ it follows that this limit is $0.$

$\square$

\textbf{Proof of Th. \ref{pippo}.} Take a function\textit{\ }$\beta\in
C^{1}(\mathbb{R})$ such that

\begin{itemize}
\item -($\beta$-i) $\forall s\in\mathbb{R},\ 0\leq\beta(s)\leq1$

\item -($\beta$-ii) $\forall s<0,\ \beta(s)>0$

\item -($\beta$-iii) $\forall s\geq0,\ \beta(s)=0$

\item -($\beta$-iiii) $\forall s\in\mathbb{R},\ \beta^{\prime}(s)<0.$ $_{.}$
\end{itemize}

Now we define a $C^{1}$-functional on $H_{r}^{1}$ as follows:
\[
F(u)=\frac{1}{2}\int\left\vert \nabla u\right\vert ^{2}dx-b\beta\left(  \int
G(u)dx\right)  ;
\]
here $b$ is a positive constant defined by
\[
b=\frac{\frac{1}{2}\int\left\vert \nabla\breve{u}\right\vert ^{2}dx+1}%
{\beta\left(  \int G(\breve{u})dx\right)  }%
\]
and $\breve{u}$ is defined by lemma \ref{ucap}. This choice of $b,$ implies
that
\[
F(\breve{u})=-1
\]
Clearly $F\ $is a $C^{1}$ functional defined on $H_{r}^{1}.$

Since $\beta$ is bounded, then $F(u)$ is bounded below and its infimum is a
number less or equal to $-1$. Then there exists a minimizing sequence $u_{n}.
$ Also, $\beta$ bounded implies $\frac{1}{2}\int\left\vert \nabla
u_{n}\right\vert ^{2}dx$ bounded and hence $u_{n}\rightarrow w$ weakly in
$\mathcal{D}_{r}^{1,2}\ $\footnote{$\mathcal{D}_{r}^{1,2}$ denotes the closure
of the set of radially simmetric $C^{\infty}$-functions with compact support
with respect to the norm $\sqrt{\int|\nabla u|^{2}dx}$.}$.$ Since the infimum
is negative we have that $\beta\left(  \int G(u_{n})dx\right)  <0,$ and hence,
by ($\beta$-iii),
\begin{equation}
\int G(u_{n})dx<0. \label{aaa2}%
\end{equation}

By (G-ii), we have that there is a constant $c_{5}>0$ such that, for $s\geq0 $%
\begin{align*}
G(s)  &  \geq\int_{0}^{s}G^{\prime}(t)dt\geq\int_{0}^{s}\left(  c_{1}%
t-c_{2}t^{p-1}\right)  dt\\
&  \geq\frac{1}{2}c_{1}s^{2}-\frac{c_{2}}{p}s^{p}\geq\frac{1}{2}c_{4}%
s^{2}-c_{5}s^{2^{\ast}}%
\end{align*}

By this inequality and (\ref{aaa2}), we get
\begin{align*}
0  &  >\int G(u_{n})dx\geq\frac{1}{2}c_{4}\int\left\vert u_{n}\right\vert
^{2}dx-c_{5}\int\left\vert u_{n}\right\vert ^{2^{\ast}}dx\\
&  \geq\frac{1}{2}c_{4}\left\Vert u_{n}\right\Vert _{L^{2}}^{2}-c_{6}%
\left\Vert \nabla u_{n}\right\Vert _{L^{2}}^{2^{\ast}}%
\end{align*}
Thus since $\left\Vert \nabla u_{n}\right\Vert _{L^{2}}$ is bounded, by the
above inequality, also $\left\Vert u_{n}\right\Vert _{L^{2}}$ is bounded and
hence also $\left\Vert u_{n}\right\Vert _{H^{1}}$ is bounded. So, by theorem
\ref{strauss}, $u_{n}\rightarrow w$ in $L^{p}$ strong, where $p$ is defined by
(G-ii). By our assumptions, the functional%
\[
u\rightarrow\int G(u)dx
\]
is continuous in $L^{p}.$ Then we have that $\int G(u_{n})dx\rightarrow\int
G(w)dx$ and hence $\beta\left(  \int G(u_{n})dx\right)  \rightarrow
\beta\left(  \int G(w)dx\right)  .$ Since $\frac{1}{2}\int\left\vert \nabla
u\right\vert ^{2}dx$ is l.s.c. it follows that $w$ is the minimum of $F.$ Thus
$w$ is different from $0$ and it satisfies the following equation:
\[
-\Delta w+\lambda G^{\prime}(w)=0
\]
where
\[
\lambda=-b\beta^{\prime}\left(  \int G(w)dx\right)
\]
By ($\beta$-iiii), we have that $\lambda\geq0,$ and since $w\neq0,$ we have
that $\lambda>0.$ Now set
\[
u(x)=w\left(  \frac{x}{\sqrt{\lambda}}\right)
\]
Clearly $u$ satisfies equation (\ref{gianna})

$\square$

\subsection{Standing waves and travelling waves\label{swtw}}

In any complex valued field theory, the simplest possible solitary waves are
the \textit{stationary waves}, namely finite energy solution having the
following form
\begin{equation}
\psi_{0}(t,x)=u(x)e^{-i\omega t}\text{, }u\text{ }\geq0, \label{sw}%
\end{equation}

In particular, for the nonlinear Scroedinger equation (\ref{NS}), substituting
(\ref{sw}) in eq.(\ref{NS}), we get
\begin{equation}
-\Delta u+W^{\prime}(u)=2\omega u \label{staticsh}%
\end{equation}

Now we can apply the results of section \ref{giovanna} and get the following theorem:

\begin{theorem}
\label{paperino}Assume that $W$ has the form (\ref{fodiw}) where $N$ satisfies
the following assumptions:
\begin{equation}
N^{\prime}(s)\geq-cs^{p-1}\ for\ any\ s\geq1, \label{nina0}%
\end{equation}
$\ $%
\begin{equation}
\exists s_{0}\in\mathbb{R}^{+}:\;N(s_{0})<0. \label{nina1}%
\end{equation}
then eq. (\ref{NS}) has finite energy solitary waves of the form (\ref{sw})
for every frequency $\omega\in\left(  E_{1},E_{0}\right)  $ where $E_{0}$ is
defined by (\ref{fodiw}) and
\[
E_{1}=\inf\left\{  a\in\mathbb{R}:\mathbb{\exists}s\in\mathbb{R}^{+}%
,\;as^{2}>E_{0}s^{2}+N(s)\right\}
\]

\end{theorem}

Notice that, by virtue of (\ref{nina1}), $E_{1}<E_{0}.$ Also, it is possible
that $E_{1}=-\infty;$ for example this happens if
\begin{equation}
W(s)=-\frac{1}{p}\left\vert s\right\vert ^{p},\ 2<p<2^{\ast} \label{Wpot}%
\end{equation}

\textbf{Proof. }We want to apply theorem \ref{pippo} to eq. (\ref{staticsh});
to this end we set
\[
G(s)=W(s)-\omega s^{2}=\left(  E_{0}-\omega\right)  s^{2}+N(s)
\]
It is trivial to verify that, for every $\omega\in\left(  E_{1},E_{0}\right)
,$ $G$ satisfies the assumptions of Th. \ref{pippo}.

$\square$

Now we will exploit the other symmetries of equation \ref{NS} to produce other
solutions. First of all, since \ref{NS} is invariant for translations, for any
$x_{0}\in\mathbb{R}^{N},$ the function
\begin{equation}
\psi_{x_{0}}(t,x)=u(x-x_{0})e^{-i\omega t} \label{swx}%
\end{equation}
is a standing wave concentrated around the point $x_{0}.$ The space rotations
do not produce other solutions since $u(x)$ is rotationally invariant.

Since the Lagrangian (\ref{lagr}) is invariant for the Galileo group, we can
obtain other solutions: we can produce travelling waves just applying the
transformation (\ref{gal}) to (\ref{swx})
\[
\psi_{x_{0},\mathbf{v}}(t,x)=u(x-x_{0}-\mathbf{v}t)e^{i(\mathbf{v\cdot}%
x-Et)},\ E=\frac{1}{2}\ v^{2}+\omega
\]

$\psi_{x_{0},\mathbf{v}}(t,x)$ is a solitary wave concentrated in the point
$x_{0}+\mathbf{v}t,\ $an hence it travels with velocity $\mathbf{v.}$

Finally, other solutions can be produced by the invariance (\ref{giulia}); for
$\theta\in\left[  0,2\pi\right)  ,$ we have the solutions
\begin{equation}
\psi_{x_{0},\mathbf{v,\theta}}(t,x)=u(x-x_{0}-\mathbf{v}t)e^{i(\mathbf{v\cdot
}x-Et+\theta)} \label{sgsh}%
\end{equation}
The invariance by time translations do not produce new solutions, since a time
translation on $\psi_{x_{0},\mathbf{v,\theta}}$ produces a space and phase
translation. Concluding, if we fix a charge $\sigma$, we obtain a radially
symmetric solution of the form (\ref{sw}); by the invariance of the equation,
this solution produces a 7-parameters family of solutions given by \ref{sgsh}.

If we consider \ref{NS} with the usual physical constants $m$ and $\hbar$,
namely equation (\ref{NSCON}) with $V=0,$ (\ref{sgsh}) takes the form
\begin{equation}
\psi_{x_{0},\mathbf{v,\theta}}(t,x)=u(x-x_{0}-\mathbf{v}t)e^{i(\mathbf{p\cdot
}x-Et)/\hbar+\theta}%
\end{equation}
where
\begin{align*}
\mathbf{p}  &  =m\mathbf{v}\\
E  &  =\frac{1}{2}mv^{2}+\omega\hbar
\end{align*}
The meaning of these relations within the swarm interpretation is the
following: $\psi_{x_{0},\mathbf{v,\theta}}(t,x)$ is interpreted as a swarm of
particles of mass $m;$ $\mathbf{p}$ is the momentum of each particle and $E$
is its energy: $\frac{1}{2}mv^{2}$ is its kinetic energy and $\omega\hbar$ is
related to the nonlinear interaction of the particles.

\subsection{Existence of solitons}

Theorem \ref{paperino} provides sufficient conditions for the existence of
solitary waves. In order to prove the existence of hylomorphic solitons (cf.
Def. \ref{hys}), first of all it is necessary to assume that the Cauchy
problem for \ref{NS} is well posed namely, that it has a unique global
solution which depends continuously on the initial data. For example, this is
the case if $W^{\prime}$ is a globally Liptschitz function; we refer to the
books \cite{Ka89} and \cite{Ca03} for more general conditions.

Moreover, it is necessary to investigate under which assumptions the energy
\begin{align*}
\mathcal{E}  &  =\int\left(  \frac{1}{2}\left\vert \nabla\psi\right\vert
^{2}+W(\psi)\right)  dx\\
&  =\int\left(  \frac{1}{2}\left\vert \nabla u\right\vert ^{2}+\frac{1}%
{2}\left\vert \nabla S\right\vert ^{2}u^{2}+W(u)\right)  dx
\end{align*}
achieves the minimum on the manifold%
\[
\mathfrak{M}_{\sigma}=\left\{  \psi\in H^{1}(\mathbb{R}^{N},\mathbb{C}%
):\int\left\vert \psi\right\vert ^{2}dx=\sigma\right\}
\]

Clearly, if $u$ minimizes the functional%
\[
J(u)=\int\frac{1}{2}|\nabla u|^{2}+N(u)dx
\]
on the manifold
\[
M_{\sigma}=\left\{  u\in H^{1}(\mathbb{R}^{N}):\int\left\vert u\right\vert
^{2}dx=\sigma\right\}
\]
then the set of all minimizers of $\mathcal{E}$ is given by $\psi
(x)=u(x)e^{i\theta},$ $\theta\in\left[  0,2\pi\right)  $

We make the following assumptions on $N:$
\begin{equation}
|N^{\prime}(s)|\leq c_{1}|s|^{q-1}+c_{2}|s|^{p-1}\text{ for some }2<q\leq
p<2^{\ast}. \label{Fp}%
\end{equation}%
\begin{equation}
N(s)\geq-c_{1}s^{2}-c_{2}|s|^{\gamma}\text{ for some }c_{1},c_{2}%
\geq0,\ {\ \gamma<2+\frac{4}{N}} \label{F0}%
\end{equation}

\begin{theorem}
\label{mainradiale} Let $N$ satisfy (\ref{Fp}), (\ref{F0}) and (\ref{nina1}).
Then, $\exists\ \bar{\sigma}$ such that, $\forall\ \sigma>\bar{\sigma}, $ $J$
has a minimizer $\bar{u}$ on $M_{\sigma}\ $which is positive and radially
symmetric around some point. Moreover, if the Cauchy problem is well posed
$\psi:=ue^{i\theta}$ is a hylomorphic soliton.
\end{theorem}

In order to have stronger results, we can replace (\ref{nina1}) with the
following hypothesis%

\begin{equation}
N(s)<-s^{\beta},\ 2<\beta<2+\frac{4}{N}\text{ for small }s. \label{F2}%
\end{equation}

In this case we find the following results concerning the existence of the
minimizer of $J(u)$ for any $\sigma$.

\begin{theorem}
\label{cor1} If (\ref{Fp}), (\ref{F0}) and (\ref{F2}) hold, then the same
conclusions of Th.\ref{mainradiale} hold for every $\sigma>0.$
\end{theorem}

In particular, for $N=3$ we have

\begin{cor}
\label{cor2} Let $N=3$. If (\ref{Fp}) and (\ref{F0}) hold and $N\in C^{3}$,
with $N^{\prime\prime\prime}(0)<0$, then the same conclusions of
Th.\ref{mainradiale} hold for every $\sigma>0.$
\end{cor}

\bigskip

For the proofs of Th.\ref{mainradiale}, Th.\ref{cor1} and Th.\ref{cor2}, we
refer to \cite{BBGM}. Analogous results have been obtained by Cazenave and
Lions \cite{CL82} when $W(u)=\frac{1}{p}|u|^{p}.$

\subsection{Dynamics of solitons\label{dynshr}}

In this section we will describe a recent result relative to the dynamics of
solitons when a potential $V(x)$ is present. As we will see, we will introduce
three parameter, $h,$ $\alpha$ and $\gamma$ in our equation and we obtain a
meaningful result for particular values of these parameters. First of all, let
us consider the following "unperturbed" Cauchy problem:
\begin{equation}
ih\frac{\partial\psi}{\partial t}=-\frac{h^{2}}{2}\Delta\psi+\frac
{1}{2h^{\alpha}}W^{\prime}\left(  h^{\gamma}\psi\right)  \label{ch}%
\end{equation}%
\begin{equation}
\psi\left(  0,x\right)  =\frac{1}{h^{\gamma}}U\left(  \frac{x-q_{0}}{h^{\beta
}}\right)  e^{\frac{i}{h}\mathbf{v}\cdot x} \label{id}%
\end{equation}
where
\begin{equation}
\beta=1+\frac{\alpha-\gamma}{2} \label{relfond}%
\end{equation}
and $U:\mathbb{R}^{N}\rightarrow\mathbb{R}$, $N\geq2$, is a positive, radially
symmetric solution of the equation
\begin{equation}
-\Delta U+W^{\prime}(U)=2\omega U \label{eq}%
\end{equation}
with
\begin{equation}
\left\Vert U\right\Vert _{L^{2}}=\sigma\label{sig}%
\end{equation}

We set
\[
u_{h}(x)=h^{-\gamma}U\left(  \frac{x}{h^{\beta}}\right)
\]
and establish a relation between $\alpha,\beta$ and $\gamma$ in order to have
stationary solution of (\ref{ch}) of the form $\psi(t,x)=u_{h}(x)e^{-i\frac
{\omega_{h}}{h}t}$: replacing this expression in (\ref{ch}) we get
\begin{equation}
-h^{2}\Delta u_{h}+\frac{1}{h^{\alpha}}W^{\prime}(h^{\gamma}u_{h})=2\omega
_{h}u_{h}. \label{calla}%
\end{equation}
If we take the explicit expression of $u_{h}$, we get
\[
-h^{2-\gamma}\Delta\left[  U\left(  \frac{x}{h^{\beta}}\right)  \right]
+\frac{1}{h^{\alpha}}W^{\prime}\left(  U\left(  \frac{x}{h^{\beta}}\right)
\right)  =2\omega_{h}h^{-\gamma}U\left(  \frac{x}{h^{\beta}}\right)
\]
and hence, by rescaling the variable $x$,
\[
-h^{2-\gamma-2\beta+\alpha}\Delta U(x)+W^{\prime}\left(  U(x)\right)
=2\omega_{h}h^{\alpha-\gamma}U\left(  x\right)  .
\]
Thus, comparing the above expression with (\ref{eq}), we get (\ref{relfond})
and%
\begin{equation}
\omega_{h}=\frac{\omega}{h^{\alpha-\gamma}}.
\end{equation}

Using the arguments of section \ref{swtw}, we have that the solution of
(\ref{ch}),(\ref{id}) is given by
\begin{equation}
\psi\left(  t,x\right)  =\frac{1}{h^{\gamma}}U\left(  \frac{x-q_{0}%
-\mathbf{v}t}{h^{\beta}}\right)  e^{\frac{i}{h}\left(  \mathbf{v}\cdot
x-Et\right)  } \label{soli}%
\end{equation}
with%
\[
E=\frac{1}{2}\mathbf{v}^{2}+\frac{\omega}{h^{\alpha-\gamma}}%
\]
Moreover if the problem (\ref{ch}),(\ref{id}) is well posed this is the unique solution.

We can interpret this result saying that the \emph{barycenter} $q(t)$ of the
solution of (\ref{ch}), (\ref{id}) (defined by \ref{bary}) satisfies the
Cauchy problem
\[
\left\{
\begin{array}
[c]{c}%
\ddot{q}=0\\
q(0)=q_{0}\\
\dot{q}(0)=\mathbf{v}%
\end{array}
\right.
\]

Let us see what happens if the problem (\ref{ch}), (\ref{id}) is perturbed,
namely let us investigate the problem
\begin{equation}
\left\{
\begin{array}
[c]{l}%
\displaystyle ih\frac{\partial\psi}{\partial t}=-\frac{h^{2}}{2}\Delta
\psi+\frac{1}{2h^{\alpha}}W^{\prime}(h^{\gamma}\psi)+V(x)\psi\\
\\
\displaystyle\psi\left(  0,x\right)  =\varphi_{h}(x)
\end{array}
\right.  \label{schr}%
\end{equation}
where
\begin{equation}
\varphi_{h}(x)=\left[  \frac{1}{h^{\gamma}}(U+w_{0})\left(  \frac{x-q_{0}%
}{h^{\beta}}\right)  \right]  e^{\frac{i}{h}\mathbf{v}\cdot x} \label{phih}%
\end{equation}
and $w_{0}$ is small, namely there is a constant $C$ such that
\begin{align*}
\left\Vert w_{0}\right\Vert _{H^{1}}  &  \leq Ch^{\alpha-\gamma};\\
\int_{\mathbb{R}^{N}}V(x)|w_{0}(x)|^{2}dx  &  \leq Ch^{\alpha-\gamma}.
\end{align*}
Also we assume that
\[
\left\Vert U+w_{0}\right\Vert _{L^{2}}=\left\Vert U\right\Vert _{L^{2}}=\sigma
\]

\begin{theorem}
\label{teo1}Let all the assumptions of Th. \ref{mainradiale} hold and let
$\sigma>\bar{\sigma}\ $($\sigma$ is defined by (\ref{sig})). Moreover we
assume that $V:V:\mathbb{R}^{N}\rightarrow\mathbb{R}$ is a $C^{2}$ function
such that:
\begin{equation}
V(x)\geq0; \tag{$V_0$}\label{V0}%
\end{equation}%
\begin{equation}
|\nabla V(x)|\leq V(x)^{b}\text{ for }|x|>R_{1}>1,b\in(0,1); \tag{$%
V_1$}\label{Vinf*}%
\end{equation}%
\begin{equation}
V(x)\geq|x|^{a}\text{ for }|x|>R_{1}>1,a>1. \tag{$V_2$}\label{Vinf1}%
\end{equation}
Finally, we assume that
\begin{equation}
\alpha>\gamma\label{ag}%
\end{equation}

Then the barycenter $q_{h}(t)$ of the solution of the problem (\ref{schr})
satisfies the following Cauchy problem:
\[
\left\{
\begin{array}
[c]{l}%
\ddot{q}_{h}(t)+\nabla V(q_{h}(t))=H_{h}(t)\\
q_{h}(0)=q_{0}\\
\dot{q}_{h}(0)=\mathbf{v}%
\end{array}
\right.
\]
where
\[
\sup_{t\in\mathbb{R}}|H_{h}(t)|\rightarrow0\;\;\text{as}\;\;h\rightarrow0
\]

\end{theorem}

\textbf{Proof: }The proof of this theorem can be found in \cite{BGM09}.

$\square$

\bigskip

\begin{rem}
The assumption (\ref{Vinf1}) is necessary if we want to identify the position
of the soliton with the barycenter (\ref{bary}). Let us see why. Consider a
soliton $\psi(x)$ and a perturbation
\[
\psi_{d}(x)=\psi(x)+\varphi\left(  x-d\right)  ,\ d\in\mathbb{R}^{N}%
\]
Even if $\varphi\left(  x\right)  \ll\psi(x),$ when $d$ is very large, the
\textquotedblleft position" of $\psi(x)$ and the barycenter of $\psi_{d}(x) $
are far from each other. In \cite{BGM09} (lemma 25), it has been proved that
this situation cannot occur provided that (\ref{Vinf1}) hold.
\end{rem}

\begin{rem}
We will give a rough explanation of the meaning of the assumption
$\alpha>\gamma$ which, in this approach to the problem, is crucial. The
energy, using eq. ($\ref{Schenergy}$) can be divided in two components: the
\textit{internal energy}
\begin{equation}
J_{h}(u)=\int\left(  \frac{h^{2}}{2}\left\vert \nabla u\right\vert ^{2}%
+W_{h}(u)\right)  dx \label{j}%
\end{equation}
and the \textit{dynamical energy}
\begin{equation}
G(u,S)=\int\left(  \frac{1}{2}\left\vert \nabla S\right\vert ^{2}+V(x)\right)
u^{2}dx \label{g}%
\end{equation}
which is composed by the \textit{kinetic energy} $\frac{1}{2}\int\left\vert
\nabla S\right\vert ^{2}u^{2}dx$ and the \textit{potential energy} $\int
V(x)u^{2}dx$. By our assumptions, the internal energy is bounded from below
and the dynamical energy is positive. As $h\rightarrow0,$ we have that
\[
J_{h}\left(  \psi_{h}\right)  \cong h^{N\beta-\alpha-\gamma}%
\]
and%
\[
G\left(  \psi_{h}\right)  \cong||\psi_{h}||_{L^{2}}^{2}\cong h^{N\beta
-2\gamma}%
\]
Then, we have that
\[
\frac{G\left(  \psi_{h}\right)  }{J_{h}\left(  \psi_{h}\right)  }\cong
h^{\alpha-\gamma}%
\]
So the assumption $\alpha-\gamma>0$ implies that, for $h\ll1,$ $G\left(
\psi_{h}\right)  \ll J_{h}\left(  \psi_{h}\right)  $, namely the internal
energy is bigger than the dynamical energy. This is the fact that guarantees
the existence and the stability of the travelling soliton for any time.
\end{rem}

\bigskip

We end this section with an heuristic proof of Th.\ref{teo1}. This proof is
not at all rigorous, but it helps to understand the underlying Dynamics. As in
section \ref{sins}, we interpret $\rho_{\mathcal{H}}=u^{2}$ as the density of
particles; then
\[
\mathcal{H=}\int\rho_{\mathcal{H}}dx
\]
is the total number of particles. By (\ref{hjqS}), each of these particle
moves as a classical particle of mass $m=1$ and hence, we can apply to the
laws of classical dynamics. In particular the center of mass defined in
(\ref{bary}) takes the following form:
\begin{equation}
q(t)=\frac{\int xm\rho_{\mathcal{H}}dx}{\int m\rho_{\mathcal{H}}dx}=\frac{\int
x\rho_{\mathcal{H}}dx}{\int\rho_{\mathcal{H}}dx}. \label{aiace}%
\end{equation}
The motion of the barycenter is not affected by the interaction between
particles (namely by the term (\ref{Qu})), but only by the external forces,
namely by $\nabla V.$ The global external force acting on the swarm of
particles is given by
\begin{equation}
\overrightarrow{F}=-\int\nabla V(x)\rho_{\mathcal{H}}dx. \label{ulisse}%
\end{equation}
Thus the motion of the center of mass $q$ follows the Newton law
\begin{equation}
\overrightarrow{F}=M{\ddot{q}}, \label{tersite}%
\end{equation}
where $M=\int m\rho_{\mathcal{H}}dx$ is the total mass of the swarm; thus by
(\ref{aiace}), (\ref{ulisse}) and (\ref{tersite}), we get
\[
{\ddot{q}}(t)=-\frac{\int\nabla V\rho_{\mathcal{H}}dx}{m\int\rho_{\mathcal{H}%
}dx}=-\frac{\int\nabla Vu^{2}dx}{m\int u^{2}dx}.
\]

If we assume that the $u(t,x)$ and hence $\rho_{\mathcal{H}}(t,x)$ is
concentrated in the point $q(t),$ we have that
\[
\int\nabla Vu^{2}dx\cong\nabla V\left(  q(t)\right)  \int u^{2}dx
\]
and so, we get
\[
m{\ddot{q}}(t)\cong-\nabla V\left(  q(t)\right)  .
\]

Notice that the equation $m{\ddot{q}}(t)=-\nabla V\left(  q(t)\right)  $ is
the Newtonian form of the Hamilton-Jacobi equation (\ref{hjS}).

\section{The nonlinear Klein-Gordon equation\label{SKG}}

\subsection{General features of NKG}

The D'Alambert equation,
\[
\square\psi=0
\]
is the simplest equation invariant for the Poincar\'{e} group, moreover it is
invariant for the "gauge" tranformation
\[
\psi\mapsto\psi+c
\]
Also, if $\psi$ is complex valued, it is invariant for the action
(\ref{giulia}). Thus, it satisfy assumptions \textbf{A-1}, \textbf{A-2} and
\textbf{A-3}, but it is linear and it does not produce solitary waves. There
exist only non-dispersive waves. Let us add to (\ref{semplice2}) a nonlinear
term:
\begin{equation}
\mathcal{L}=\frac{1}{2}\left\vert \partial_{t}\psi\right\vert ^{2}-\frac{1}%
{2}\left\vert \nabla\psi\right\vert ^{2}-W(\psi) \label{lagra}%
\end{equation}
where
\[
W:\mathbb{C}\rightarrow\mathbb{R}%
\]
satisfies the following assumption,
\[
W\left(  e^{i\theta}\psi\right)  =W\left(  \psi\right)
\]
namely $W\left(  \psi\right)  =F(\left\vert \psi\right\vert )$ for some
function $F=\mathbb{R}\rightarrow\mathbb{R}.$ This is simplest non-linear
Lagrangian invariant for the Poincar\'{e} group and the trivial gauge action
(\ref{giulia}).

The equation of motion relative to the Lagrangian (\ref{lagra}) is the
following:
\begin{equation}
\square\psi+W^{\prime}\left(  \psi\right)  =0 \tag{NKG}\label{KG}%
\end{equation}
where $W^{\prime}\left(  \psi\right)  $ is intended as in (\ref{w'}).

In the following sections we will see that equation \ref{KG}, with suitable
(but very general) assumptions on $W$ produces a very rich model in which
there are solitary waves and solitons. Moreover we will see that these
solitons behave as relativistic particles.

If $W^{\prime}(\psi)$ is linear, namely $W^{\prime}(\psi)=m^{2}\psi,$ then eq.
\ref{KG} reduces to the Klein-Gordon equation
\begin{equation}
\square\psi+m^{2}\psi=0. \label{KGE}%
\end{equation}
Among the solutions of the Klein-Gordon equations there are the \textit{wave
packets }which behave as solitary waves but disperse in space as time goes on.
On the contrary, if $W$ has a nonlinear suitable component, the wave packets
do not disperse and give hylomorphic solitons.

\bigskip

Sometimes, it will be useful to write $\psi$ in polar form, namely
\begin{equation}
\psi(t,x)=u(t,x)e^{iS(t,x)} \label{polar}%
\end{equation}

In this case the action $\int\mathcal{L}dxdt$ takes the fom
\begin{equation}
\mathcal{S}(u,S)=\frac{1}{2}\int\left(  \partial_{t}u^{2}\right)  -\left\vert
\nabla u\right\vert ^{2}+\left[  \left(  \partial_{t}S^{2}\right)  -\left\vert
\nabla S\right\vert ^{2}\right]  u^{2}dxdt-\int W(u)dxdt=0
\end{equation}
and equation \ref{KG} becomes:%

\begin{equation}
\square u-\left[  \left(  \partial_{t}S^{2}\right)  -\left|  \nabla S\right|
^{2}\right]  u^{2}+W^{\prime}(u)=0 \label{KG1}%
\end{equation}

\begin{equation}
\partial_{t}\left(  u^{2}\partial_{t}S\right)  -\nabla\cdot\left(  u^{2}\nabla
S\right)  =0 \label{KG2}%
\end{equation}

\subsection{First integrals of NKG and the hylenic ratio\label{24}}

Easy computations and the results of section \ref{cl}, show that the integral
of motion of \ref{KG} are given by the following expression:

\begin{itemize}
\item \textbf{Energy}. We get
\begin{equation}
\mathcal{E}=\int\left[  \frac{1}{2}\left\vert \partial_{t}\psi\right\vert
^{2}+\frac{1}{2}\left\vert \nabla\psi\right\vert ^{2}+W(\psi)\right]  dx
\label{energy}%
\end{equation}

\end{itemize}

Using (\ref{polar}) we get:
\begin{equation}
\mathcal{E}=\int\left[  \frac{1}{2}\left(  \partial_{t}u\right)  ^{2}+\frac
{1}{2}\left|  \nabla u\right|  ^{2}+\frac{1}{2}\left[  \left(  \partial
_{t}S\right)  ^{2}+\left|  \nabla S\right|  ^{2}\right]  u^{2}+W(u)\right]  dx
\label{penergy}%
\end{equation}

\begin{itemize}
\item \textbf{Momentum. }We have
\begin{equation}
\mathbf{P}=-\operatorname{Re}\int\partial_{t}\psi\overline{\nabla\psi}\;dx
\label{momentum}%
\end{equation}

Using (\ref{polar}) we get:
\begin{equation}
\mathbf{P}=-\int\left(  \partial_{t}u\,\nabla u+\partial_{t}S\,\nabla
S\;u^{2}\right)  \;dx \label{pmomentum}%
\end{equation}

\item \textbf{Angular momentum. }We have
\begin{equation}
\mathbf{M}=\operatorname{Re}\int\mathbf{x}\times\nabla\psi\overline
{\partial_{t}\psi}\;dx \label{amomentum}%
\end{equation}

Using (\ref{polar}) we get:
\begin{equation}
\mathbf{M}=\int\left(  \mathbf{x}\times\nabla S\ \partial_{t}S\ u^{2}%
+\mathbf{x}\times\nabla u\,\partial_{t}u\right)  \;dx \label{apmomentum}%
\end{equation}

\item \textbf{Hylenic Charge. }We have
\[
\mathcal{H}=\operatorname{Im}\int\partial_{t}\psi\overline{\psi}\;dx
\]

\end{itemize}

Using (\ref{polar}) we get:
\begin{equation}
\mathcal{H}=\int\partial_{t}S\,u^{2}dx \label{cha}%
\end{equation}

\begin{itemize}
\item \textbf{Ergocenter velocity. }If we take the Lagrangian (\ref{lagra}),
the quantity preserved by the Lorentz transformation, is the following
\begin{equation}
\mathbf{K}=t\mathbf{P}-\int\mathbf{x}\left[  \frac{1}{2}\left\vert
\partial_{t}\psi\right\vert ^{2}+\frac{1}{2}\left\vert \nabla\psi\right\vert
^{2}+W(\psi)\right]  dx \label{kappa}%
\end{equation}

\end{itemize}

The computation of $\mathbf{K}$ is more involved than the prrevious ones and
we will make it in details:

\textbf{Proof. }Let us compute $K_{i}$ using Th. \ref{noe}; in this case the
parameter $\lambda$ is the velocity $v_{i}$ which appears in (\ref{LT}); we
have%
\begin{align*}
\rho_{K_{i}}  &  =\operatorname{Re}\left(  \frac{\partial\mathcal{L}}%
{\partial\psi_{t}}\overline{\frac{\partial\psi}{\partial v_{i}}}\right)
-\mathcal{L}\frac{\partial t}{\partial v_{i}}\\
&  =\operatorname{Re}\left(  \partial_{t}\psi\overline{\left[  \frac
{\partial\psi}{\partial t}\frac{\partial t}{\partial v_{i}}+%
{\displaystyle\sum\limits_{k=1}^{3}}
\frac{\partial\psi}{\partial x_{k}}\frac{\partial x_{k}}{\partial v_{i}%
}\right]  }\right)  -\left(  \frac{1}{2}\left\vert \partial_{t}\psi\right\vert
^{2}-\frac{1}{2}\left\vert \nabla\psi\right\vert ^{2}-W(\psi)\right)
\frac{\partial t}{\partial v_{i}}%
\end{align*}
where the derivative with respect to $v_{i}$ need to be computed for
$v_{i}=0.$

Since for $k\neq i,\ \frac{\partial x_{k}}{\partial v_{i}}=0,$ we have that
\begin{align*}
\rho_{K_{i}}  &  =\left\vert \partial_{t}\psi\right\vert ^{2}\frac{\partial
t}{\partial v_{i}}+\operatorname{Re}\left(  \partial_{t}\psi\overline
{\partial_{x_{i}}\psi}\right)  \cdot\frac{\partial x_{i}}{\partial v_{i}%
}-\left(  \frac{1}{2}\left\vert \partial_{t}\psi\right\vert ^{2}-\frac{1}%
{2}\left\vert \nabla\psi\right\vert ^{2}-W(\psi)\right)  \frac{\partial
t}{\partial v_{i}}\\
&  =\left(  \frac{1}{2}\left\vert \partial_{t}\psi\right\vert ^{2}+\frac{1}%
{2}\left\vert \nabla\psi\right\vert ^{2}+W(\psi)\right)  \frac{\partial
t}{\partial v_{i}}+\operatorname{Re}\left(  \partial_{t}\psi\overline
{\partial_{x_{i}}\psi}\right)  \cdot\frac{\partial x_{i}}{\partial v_{i}}\\
&  =\rho_{\mathcal{E}}\frac{\partial t}{\partial v_{i}}-\rho_{P_{i}}\cdot
\frac{\partial x_{i}}{\partial v_{i}}%
\end{align*}

Also we have
\begin{align*}
\left(  \frac{\partial t}{\partial v_{i}}\right)  _{v_{i}=0}  &  =\left(
\frac{\partial}{\partial v_{i}}\frac{t-v_{i}x}{\sqrt{1-v_{i}^{2}}}\right)
_{v_{i}=0}=-x\\
\left(  \frac{\partial x}{\partial v_{i}}\right)  _{v_{i}=0}  &  =\left(
\frac{\partial}{\partial v_{i}}\frac{x-v_{i}t}{\sqrt{1-v_{i}^{2}}}\right)
_{v_{i}=0}=-t
\end{align*}
so that%
\[
\rho_{K_{i}}=-\rho_{\mathcal{E}}x_{i}+\rho_{P_{i}}t
\]
Integrating in the space we get the $i$-th component of (\ref{kappa}).

$\square$

\bigskip

Let us interpret (\ref{kappa}) in a more meaningful way. If we derive the
terms of (\ref{kappa}) with respect to $t$, we get
\begin{equation}
\mathbf{P=}\frac{d}{dt}\int\mathbf{x}\left[  \frac{1}{2}\left\vert
\partial_{t}\psi\right\vert ^{2}+\frac{1}{2}\left\vert \nabla\psi\right\vert
^{2}+W(\psi)\right]  dx=\int\mathbf{x}\rho_{\mathcal{E}}dx \label{elle1}%
\end{equation}

Now, we define the \emph{ergocenter} (or \textit{barycenter) }as follows
\begin{equation}
\mathbf{Q:=}\frac{\int\mathbf{x}\left[  \frac{1}{2}\left\vert \partial_{t}%
\psi\right\vert ^{2}+\frac{1}{2}\left\vert \nabla\psi\right\vert ^{2}%
+W(\psi)\right]  dx}{\int\left[  \frac{1}{2}\left\vert \partial_{t}%
\psi\right\vert ^{2}+\frac{1}{2}\left\vert \nabla\psi\right\vert ^{2}%
+W(\psi)\right]  dx}=\frac{\int\mathbf{x}\rho_{\mathcal{E}}dx}{\mathcal{E}};
\label{dergo}%
\end{equation}
then, by the conservation of $\mathcal{E}$ and eq.(\ref{elle1}), we get
\begin{equation}
\mathbf{\dot{Q}=}\frac{\mathbf{P}}{\mathcal{E}} \label{ergo}%
\end{equation}

Concluding, the Poincar\'{e} group provides 10 independent integral of motions
which are $\mathcal{E},$ $\mathbf{P,}$ $\mathbf{M,}$ $\mathbf{K;}$ they can be
replaced by integral of motions $\mathcal{E},$ $\mathbf{P,}$ $\mathbf{M,}$
$\mathbf{\dot{Q}}$ since also these quantities are independent.

Notice the difference between (\ref{leann}) and (\ref{kappa}) and consequently
the difference between the \textit{hylecenter} (\ref{bary}) and the
\textit{ergocenter }(\ref{ergo}). In \ref{NS} the barycenter or "center of
mass" or coincide with the \textit{hylecenter or }"center of hylenic charge";
in \ref{KG}, the barycenter coincide with the \textit{ergocenter or }"center
of energy".

The precise definition of mass and its meaning will be discussed at pag.
\pageref{ma}.

\bigskip

We now assume that $W$ is of class $C^{2}$ and we set
\begin{equation}
W(s)=\frac{1}{2}m^{2}s^{2}+N(s) \label{wn}%
\end{equation}
where $m^{2}=W^{\prime\prime}(0).$

\begin{theorem}
\label{hr}If $W$ is of class $C^{2}$, then we have that
\[
E_{0}:=\,\underset{\varepsilon\rightarrow0}{\lim}\;\underset{\Psi\in
X_{\varepsilon}}{\inf}\frac{\mathcal{E}\left(  \Psi\right)  }{\left|
\mathcal{H}\left(  \Psi\right)  \right|  }=m
\]

\end{theorem}

\textbf{Proof. }We have $\Psi=(\psi,\psi_{t})\equiv\left(  u,u_{t}%
,\omega,\mathbf{k}\right)  ;$ then by (\ref{penergy}), (\ref{cha}) and
(\ref{NN})
\begin{align*}
\frac{\mathcal{E}\left(  \Psi\right)  }{\left\vert \mathcal{H}\left(
\Psi\right)  \right\vert }  &  =\frac{\int\left[  \frac{1}{2}\left(
\partial_{t}u\right)  ^{2}+\frac{1}{2}\left\vert \nabla u\right\vert
^{2}+\frac{1}{2}\left[  (\partial_{t}S)^{2}+(\nabla S)^{2}\right]
u^{2}+W(u)\right]  dx}{\left\vert \int\omega\,u^{2}dx\right\vert }\\
&  \geq\frac{\int\left[  \frac{1}{2}\omega^{2}u^{2}+\frac{1}{2}m^{2}%
u^{2}+N(u)\right]  dx}{\int\left\vert \omega\right\vert \,u^{2}dx}%
\end{align*}
Since
\begin{align*}
\int\left\vert \omega\right\vert \,u^{2}dx  &  \leq\left(  \int\omega
^{2}\,u^{2}dx\right)  ^{1/2}\cdot\left(  \int\,u^{2}dx\right)  ^{1/2}\\
&  =\frac{1}{m}\left(  \int\omega^{2}\,u^{2}dx\right)  ^{1/2}\cdot\left(
\int\,m^{2}u^{2}dx\right)  ^{1/2}\\
&  \leq\frac{1}{2m}\left[  \int\omega^{2}u^{2}dx+\int\,m^{2}u^{2}dx\right] \\
&  =\frac{1}{2m}\int\left(  \omega^{2}+m^{2}\right)  u^{2}dx
\end{align*}
we have that
\[
\frac{\mathcal{E}\left(  \Psi\right)  }{\left\vert \mathcal{H}\left(
\Psi\right)  \right\vert }\geq\frac{\int\left[  \frac{1}{2}\omega^{2}%
u^{2}+\frac{1}{2}m^{2}u^{2}+N(u)\right]  dx}{\frac{1}{2m}\int\left(
\omega^{2}+m^{2}\right)  u^{2}dx}=m+\frac{\int N(u)dx}{\frac{1}{2m}\int\left(
\omega^{2}+m^{2}\right)  u^{2}dx}%
\]
Then since $N(u)=O(u^{3})$ for $u\rightarrow0,$ we have that
\[
\underset{\varepsilon\rightarrow0}{\lim}\;\underset{\Psi\in X_{\varepsilon}%
}{\inf}\frac{\mathcal{E}\left(  \Psi\right)  }{\left\vert \mathcal{H}\left(
\Psi\right)  \right\vert }\geq m
\]

In order to prove the opposite inequality, take $\Psi_{\delta,R}=(\delta
u_{R},-i\delta u_{R})\equiv\left(  u,0,1,\mathbf{0}\right)  $\ where
\begin{equation}
u_{R}(x)=\left\{
\begin{array}
[c]{cc}%
1 & if\;\;|x|<R\\
0 & if\;\;|x|>R+1\\
1+R-|x| & if\;\;R<|x|<R+1
\end{array}
\right.  \label{uR}%
\end{equation}

Then
\begin{align*}
\underset{\Psi\in X_{\varepsilon}}{\inf}\frac{\mathcal{E}\left(  \Psi\right)
}{\left\vert \mathcal{H}\left(  \Psi\right)  \right\vert }  &  \leq
\frac{\mathcal{E}\left(  \Psi_{\varepsilon,R}\right)  }{\left\vert
\mathcal{H}\left(  \Psi_{\varepsilon,R}\right)  \right\vert }=\frac
{\varepsilon^{2}\int\left[  \frac{1}{2}\left\vert \nabla u_{R}\right\vert
^{2}+\frac{1}{2}u_{R}^{2}+\frac{1}{2\varepsilon^{2}}W(\varepsilon u)\right]
dx}{\varepsilon^{2}\int\,u_{R}^{2}dx}\\
&  =\frac{\int\left[  \frac{1}{2}\left\vert \nabla u_{R}\right\vert ^{2}%
+u_{R}^{2}+\frac{1}{2\varepsilon^{2}}N(\varepsilon u)\right]  dx}{\int
\,u_{R}^{2}dx}\leq1+\frac{1}{2}\frac{\int\left\vert \nabla u_{R}\right\vert
^{2}dx}{\int\,u_{R}^{2}dx}+\frac{\int\frac{1}{2\varepsilon^{2}}N(\varepsilon
u)dx}{\int\,u_{R}^{2}dx}\\
&  =1+O\left(  \frac{1}{R}\right)  +O\left(  \varepsilon\right)
\end{align*}
$\square$

\subsection{Swarm interpretation of NKG\label{sinkg}}

Before giving the swarm interpretation to equation \ref{KG}, we will write it
with the usual physical constants $c,\ m$ and $\hslash:$%
\begin{equation}
\frac{\partial^{2}\psi}{\partial t^{2}}-c^{2}\Delta\psi+W^{\prime}(\psi)=0
\label{uh}%
\end{equation}
with
\[
W(u)=\frac{m^{2}c^{4}}{2\hslash^{2}}u^{2}+N(u)
\]
Here $c\ $has the dimension of a velocity (and represents the speed of light),
$m$ has the dimension of \textit{mass} and $\hslash$ is the Plank constant.

The polar form of $\psi$ is written as follows
\begin{equation}
\psi(t,x)=u(t,x)e^{iS(t,x)/\hslash} \label{polar3}%
\end{equation}
and equations (\ref{KG1}) and (\ref{KG2}) become%

\begin{equation}
\hslash^{2}\left(  \partial_{t}^{2}u-c^{2}\Delta u+N^{\prime}(u)\right)
+\left(  -\partial_{t}^{2}S+c^{2}\left\vert \nabla S\right\vert ^{2}%
+m^{2}c^{4}\right)  u=0 \label{KG1c}%
\end{equation}

\begin{equation}
\partial_{t}\left(  u^{2}\partial_{t}S\right)  -c^{2}\nabla\cdot\left(
u^{2}\nabla S\right)  =0 \label{KG2c}%
\end{equation}

The continuity equation (\ref{CE}) for \ref{KG} is given by (\ref{KG2c}). This
equation allows us to interprete the matter field to be a fluid composed by
particles whose density is given by
\[
\rho_{\mathcal{H}}=-u^{2}\partial_{t}S
\]
and which move in the velocity field
\begin{equation}
\mathbf{v}=-\frac{\nabla S}{c^{2}\partial_{t}S}. \label{vel}%
\end{equation}

If
\begin{equation}
\hslash^{2}\left(  \partial_{t}^{2}u-c^{2}\Delta u+N^{\prime}(u)\right)
\ll\left(  -\partial_{t}^{2}S+c^{2}\left\vert \nabla S\right\vert ^{2}%
+m^{2}c^{4}\right)  u,
\end{equation}
namely, if $\hslash$ is very small with respect to the other quantities
involved, equation (\ref{KG1c}) can be approximated by
\begin{equation}
\partial_{t}^{2}S=c^{2}\left\vert \nabla S\right\vert ^{2}+m^{2}c^{4}.
\end{equation}
or
\begin{equation}
\partial_{t}S+\sqrt{m^{2}c^{4}+c^{2}\left\vert \nabla S\right\vert ^{2}}=0
\label{hjKG}%
\end{equation}
This is the Hamilton-Jacobi equation of a free relativistic particle of rest
mass $m$ (cf. eq. \ref{tinarel1}) whose trajectory $q(t)$, by (\ref{vel})
satisfies the equation
\begin{equation}
\dot{q}=-\frac{\nabla S}{c^{2}\partial_{t}S} \label{velq}%
\end{equation}
(cf. eq. \ref{tinarel2})

If we do not assume (\ref{rosaS}), equation (\ref{hjS}) needs to be replaced
by%
\[
\partial_{t}S=\pm\sqrt{m^{2}c^{4}+c^{2}\left\vert \nabla S\right\vert
^{2}+Q(u)}%
\]
with
\[
Q(u)=\hslash^{2}\left(  \partial_{t}^{2}u-c^{2}\Delta u+N^{\prime}(u)\right)
\]

The term $Q(u)$ can be regarded as a field describing a sort of interaction
between particles.

\ Given a wave of the form (\ref{polar3}), the local frequency and the local
wave number are defined as follows:
\begin{align*}
\omega(t,x)  &  =-\frac{\partial_{t}S(t,x)}{\hslash}\\
\mathbf{k}(t,x)  &  =\frac{\nabla S(t,x)}{\hslash};
\end{align*}
the energy of each particle moving according to (\ref{hjKG}), is given by
\[
E=\partial_{t}S
\]
and its momentum is given by
\[
\mathbf{p}=\nabla S;
\]
thus we have that
\begin{align*}
E  &  =\hslash\omega\\
\mathbf{p}  &  =\hslash\mathbf{k};
\end{align*}
these two equations are the De Broglie relation.

Thus, eq. (\ref{vel}) becomes%
\begin{equation}
\mathbf{v}=\frac{\mathbf{k}}{\omega}=\frac{\mathbf{p}}{E}. \label{vel1}%
\end{equation}

\subsection{Existence of solitary waves and solitons in NKG}

The easiest way to produce solitary waves of \ref{KG} consists in solving the
static equation
\begin{equation}
-\triangle u+W^{\prime}(u)=0 \label{KGS}%
\end{equation}
and setting
\begin{equation}
\psi_{v}(t,x)=\psi_{v}(t,x_{1},x_{2},x_{3})=u\left(  \frac{x_{1}-vt}%
{\sqrt{1-v^{2}}},x_{2},x_{3}\right)  ; \label{pina}%
\end{equation}
$\psi_{v}(t,x)$ is a solution of eq. \ref{KG} which represents a bump which
travels in the $x_{1}$-direction with speed $v.$

Thus by Theorem \ref{pippo}, we obtain the following result:

\begin{theorem}
Assume that $W$ satisfies (G-i), (G-ii), (G-iii). Then eq. \ref{KG} has real
valued solitary waves of the form (\ref{pina}).
\end{theorem}

However, it would be interesting to assume
\begin{equation}
W\geq0; \label{due}%
\end{equation}
in fact the energy of a solution of equation \ref{KG} is given by
\[
E(\psi)=\int\left[  \frac{1}{2}\left\vert \partial_{t}\psi\right\vert
^{2}+\frac{1}{2}\left\vert \nabla\psi\right\vert ^{2}+W(\psi)\right]  dx
\]
Thus, (\ref{due}) implies that every state $\psi$ has positive energy. In this
case, the positivity of the energy, not only is an important request for the
physical models related to this equation, but it provides good \textit{a
priori} estimates for the solutions of the relative Cauchy problem. These
estimate allows to prove the existence and the well-posedness results under
very general assumptions on $W$. Unfortunately Derrick \cite{D64}, in a very
well known paper, has proved that request (\ref{due}) implies that equation
(\ref{KGS}) has only the trivial solution. His proof is based on the following
equality (which in a different form was also found by Pohozaev \label{poho}).
The Derrick-Pohozaev identity (see e.g. (\cite{sammomme})) states that for any
finite energy solution $u$ of eq. \ref{KG} it holds
\begin{equation}
\left(  \frac{1}{N}-\frac{1}{2}\right)  \int\left\vert \nabla u\right\vert
^{2}dx+\int W(u)dx=0 \label{ventuno}%
\end{equation}
Clearly the above inequality and (\ref{due}) imply that $u\equiv0$ for
$N\geq2$.

However, we can try to prove the existence of solitons of eq. \ref{KG} (with
assumption (\ref{due})) exploiting the possible existence of \textit{standing
waves }(as defined by\textit{\ }(\ref{sw})), since this fact is not prevented
by eq.(\ref{ventuno}).

\noindent Substituting (\ref{sw}) in eq. \ref{KG}, we get
\begin{equation}
-\Delta u+W^{\prime}(u)=\omega^{2}u \label{static}%
\end{equation}

Since the Lagrangian (\ref{lagra}) is invariant for the Lorentz group, we can
obtain other solutions $\psi_{1}(t,x)$ just making a Lorentz transformation on
it. Namely, if we take the velocity $\mathbf{v}=(v,0,0),$ $\left\vert
v\right\vert <1$, and set
\[
t^{\prime}=\gamma\left(  t-vx_{1}\right)  ,\text{ }x_{1}^{\prime}%
=\gamma\left(  x_{1}-vt\right)  ,\text{ }x_{2}^{\prime}=x_{2},\text{ }%
x_{3}^{\prime}=x_{3}\;\;\;\text{with}\;\;\;\gamma=\frac{1}{\sqrt{1-v^{2}}}%
\]
it turns out that $\psi_{1}(t,x)=\psi(t^{\prime},x^{\prime}) $ is a solution
of \ref{KG}.

In particular given a standing wave $\psi(t,x)=u(x)e^{-i\omega t},$ the
function $\psi_{\mathbf{v}}(t,x):=\psi(t^{\prime},x^{\prime})$ is a solitary
wave which travels with velocity $\mathbf{v.}$ Thus, if $u(x)=u(x_{1}%
,x_{2},x_{3})$ is any solution of Eq. (\ref{static}), then
\begin{equation}
\psi_{\mathbf{v}}(t,x_{1},x_{2},x_{3})=u\left(  \gamma\left(  x_{1}-vt\right)
,x_{2},x_{3}\right)  e^{i(\mathbf{k\cdot x}-\omega t)},\;\text{ }
\label{solitone}%
\end{equation}
is a solution of Eq. \ref{KG} provided that
\begin{equation}
\omega=\gamma\omega_{0}\;\;\text{and\ \ }\;\mathbf{k}=\gamma\omega
_{0}\mathbf{v} \label{kome}%
\end{equation}
Notice that (\ref{pina}) is a particular case of (\ref{solitone}) when
$\omega_{0}=0.$

\bigskip

We write $W$ in the form (\ref{wn}) and we make the following assumptions:

\begin{itemize}
\item (W-i) \textbf{(Positivity}) $W(s)\geq0$

\item (W-ii) \textbf{(Nondegeneracy}) $W=$ $W(s)$ ( $s\geq0)$ is $C^{2}$ near
the origin with $W(0)=W^{\prime}(0)=0;\;W^{\prime\prime}(0)=m^{2}$\ $>0$

\item (W-iii) \textbf{(Hylomorphy}) $\exists s_{0}:\;N(s_{0})<0$\ 

\item (W--iiii) \textbf{(Growth}) there is a constant $c>0$ such that
\[
N^{\prime}(s)\geq-c_{1}s-c_{2}s^{p-1},\ \ 2<p<2^{\ast}%
\]

\end{itemize}

Here there are some comments on assumptions (W-i), (W-ii), (W-iii).

(W-i) implies that the energy is positive; if this condition does not hold, it
is possible to have solitary waves, but not hylomorphic waves (cf. Proposition
16 of \cite{BF09TA}).

(W-ii) In order to have solitary waves it is necessary to have $W^{\prime
\prime}(0)\geq0.$ There are some results also when $W^{\prime\prime}(0)=0 $
(null-mass case, see e.g. \cite{BL81} and \cite{BBR07}), however the most
interesting situation occurs when $W^{\prime\prime}(0)>0.$

(W-iii) This is the crucial assumption which characterizes the potentials
which might produce hylomorphic solitons. This assumption permits to have
states $\Psi$ with hylomorphy ratio $\Lambda\left(  \Psi\right)  <m$.
Actually, $N$ is the nonlinear term which, when it is negative, produces a
attractive \textquotedblright force\textquotedblright.

We have the following result:

\begin{theorem}
\label{lillo}Assume that (W-i),...,(W-iiii) hold, then eq. \ref{KG} has finite
energy solitary waves of the form $\;\psi(t,x)=u(x)e^{-i\omega t}$ for every
frequency $\omega\in\left(  m_{0},m\right)  $ where
\[
m_{0}=\inf\left\{  a\in\mathbb{R}:\mathbb{\exists}u\in\mathbb{R}^{+}%
,\;\frac{1}{2}a^{2}u^{2}>W(u)\right\}
\]

\end{theorem}

Notice that by (W-iii), $m_{0}<m,$ then the interval $\left(  m_{0},m\right)
$ is not empty.

\textbf{Proof. }By the previous discussion, it is sufficient to show that
equation (\ref{static}) has a solution $u$ with finite energy. The solutions
of finite energy of (\ref{static}) are the critical points in the Sobolev
space $H^{1}\left(  \mathbb{R}^{3}\right)  $ of the \textit{reduced action}
functional:
\begin{equation}
J(u)=\frac{1}{2}\int\left\vert \nabla u\right\vert ^{2}dx+\int G(u)dx,\text{
}G\left(  u\right)  =W(u)-\frac{1}{2}\omega^{2}u^{2}\text{ }%
\end{equation}

Now we apply theorem \ref{pippo}.

It is easy to check that for every frequency $\omega\in\left(  m_{0},m\right)
,$ the required assumptions are satisfied.

$\square$

\bigskip

In \cite{BBBM} the existence of soliton is proved. The proof is quite involved
and will not discussed here; we only refer that it is necessary to strengthen
assumption (W-iiii) with the following one:

\begin{itemize}
\item (W-iiii')\textbf{(Growth condition}) Al least one of the following
assumptions holds:

\begin{itemize}
\item (a) there are constants $a,b>0,$ $2<p<2N/(N-2)$ such that for any $s>0:
$%
\[
|N^{\prime}(s)|\ \leq as^{p-1}+bs^{2-\frac{2}{p}}.
\]

\item (b) $\exists s_{1}>s_{0}:$ $N^{\prime}(s_{1})\geq0.$
\end{itemize}
\end{itemize}

\subsection{Dynamical properties\label{DP}}

In this section we will show that the solitons and the solitary waves relative
to eq. \ref{KG} behave as relativistic bodies. In fact, the relativistic
effects like the space contraction, the time dilation and the equality between
mass and energy\ are consequences of of the variational principle \textbf{A-1
}and the invariance for the Poincar\'{e} group.\textbf{\ }

First of all observe that, by Eq. (\ref{solitone}), the following theorem follows:

\begin{theorem}
Any moving solitary wave experiences a contraction in the direction of its
movement of a factor $1/\gamma$ with $\gamma=\frac{1}{\sqrt{1-v^{2}}}$.
\end{theorem}

Thus the space contraction is a trivial fact. On the contrary, the time
dilation needs a more subtle computation.

The standing waves of eq. can be considered as a clock. Let us denote by
$\mathbf{q}\left(  t\right)  $ the position of our clock at the time $t$.

If we assume that at $t=0,$ $\mathbf{q}(0)=(0,0,0),$ the motion of the clock
is given by
\[
\mathbf{q}(t)=(vt,0,0);
\]
then the behavior of the moving clock at the point $\mathbf{q}(t)$ is obtained
replacing $x$ by $\mathbf{q}(t)$ in Eq. (\ref{solitone}):
\begin{align*}
\psi_{v}(t,\mathbf{q}(t))  &  =\psi_{v}(t,vt,0,0)\\
&  =u\left(  0,0,0\right)  e^{i(\mathbf{k\cdot\mathbf{q}}(t)-\omega t)};
\end{align*}
taking into account eq. (\ref{kome}), we get
\begin{align*}
\mathbf{k\cdot\mathbf{q}}(t)-\omega t  &  =\gamma\omega_{0}v_{1}\cdot
v_{1}t-\omega t\\
&  =(\gamma\omega_{0}v_{1}^{2}-\gamma\omega_{0})t\\
&  =\gamma\omega_{0}\left(  v_{1}^{2}-1\right)  t\\
&  =\frac{\omega_{0}}{\gamma}t
\end{align*}
Then
\[
\psi_{v}(t,\mathbf{q}(t))=u\left(  0,0,0\right)  e^{^{-i\frac{\omega_{0}%
}{\gamma}t}}%
\]
this equation shows that our moving clock is vibrating with a frequency
\[
\frac{\omega_{0}}{\gamma}=\omega_{0}\sqrt{1-v^{2}}%
\]
Since the intervals of time measured by a clock are inversely proportional to
the frequency of the vibrations, we get that
\[
\Delta T=\frac{\Delta T_{0}}{\sqrt{1-v^{2}}}.
\]
Then we get the following

\begin{theorem}
A moving clock moves slower than a resting clock by a factor $\gamma^{-1}.$
\end{theorem}

In classical mechanics the mass $m$ can be defined as a quantity which relates
the momentum $\mathbf{P=}\left(  P_{1},P_{2},P_{3}\right)  $ to the velocity
$\mathbf{v=}\left(  v_{1},v_{2},v_{3}\right)  \;$by the following formula
\[
\mathbf{P}=m\mathbf{v}%
\]
Since the momentum of a solitary wave is defined by (\ref{mom}), it is
possible to define the mass of a solitary wave by the above formula and to
compute it.

In the case of the Schroedinger equation, the velocty of a soliton is given by
$\mathbf{\dot{q}}$ where $\mathbf{q}$ is defined by (\ref{bary}); then by
(\ref{leann3}) we have that%
\begin{equation}
mass=\frac{\mathbf{P}}{\mathbf{\dot{q}}}=\mathcal{H} \label{ma}%
\end{equation}
namely the mass of the soliton equals the hylenic charge. Actually, if we add
the usual constant as in eq. (\ref{NSCON}), we get that%
\[
mass=m\mathcal{H}%
\]
This equation is consistent with the swarm intepretation of section
\ref{sins}: it says that the soliton consists of a number $\mathcal{H}$ of
particles of mass $m.$

In the case of the equation \ref{KG}, the velocty of a soliton is given by
$\mathbf{\dot{Q}}$ where $\mathbf{Q}$ is defined by (\ref{dergo}); then by
(\ref{ergo}) we have that%
\[
mass=\frac{\mathbf{P}}{\mathbf{\dot{Q}}}=\mathcal{E}%
\]
namely the mass of the soliton equals its energy. If we add the usual constant
as in eq. (\ref{uh}), we get that%
\[
mass=\frac{\mathcal{E}}{c^{2}}%
\]
namely the celebrated Einstein equation. Namely, we have the following:

\begin{theorem}
The mass of a solitary wave of eq. \ref{KG} is proportional to its energy
with\ the factor of proportionality $c^{-2}.$
\end{theorem}


\bigskip

\begin{thebibliography}{99}                                                                                               %


\bibitem {BBBM}\textsc{J. Bellazzini, V. Benci, C. Bonanno, A.M.\ Micheletti,}
\emph{\ Solitons for the Nonlinear Klein-Gordon-Equation}, to appear. (arXiv:0712.1103)

\bibitem {BBR07}\textsc{M.Badiale, V.Benci, S.Rolando,} \emph{A nonlinear
elliptic equation with singular potential and applications to nonlinear field
equations}, J. Eur. Math. Soc., \textbf{9} (2007), 355--381

\bibitem {hylo}\textsc{J. Bellazzini, V. Benci, C. Bonanno, E. Sinibaldi,}
\emph{\ }\textit{Hylomorphic solitons in the nonlinear Klein-Gordon
equation,,} to appear. (arXiv:0810.5079)

\bibitem {BBGM}\textsc{J. Bellazzini, V. Benci, M. Ghimenti, A.M. Micheletti,}
\emph{\ On the existence of the fundamental eigenvalue of an elliptic problem
in $\mathbb{R}^{N}$ }, Adv. Nonlinear Stud. \textbf{7} (2007), 439--458

\bibitem {bebo}\textsc{Bellazzini J., Bonanno C., }\emph{Nonlinear
Schr\"{o}dinger equations with strongly singular potentials. }preprint.

\bibitem {BF02}\textsc{Benci V., Fortunato D.,}\ \textit{Solitary waves of the
nonlinear Klein-Gordon field equation coupled with the Maxwell equations,
}Rev. Math. Phys. \textbf{14} (2002), 409-420.

\bibitem {sammomme}\textsc{Benci V. Fortunato D.,}\textit{\ Solitary waves in
classical field theory, }in Nonlinear Analysis and Applications to Physical
Sciences, V. Benci A. Masiello Eds \ Springer, Milano (2004), 1-50.

\bibitem {befov07}\textsc{Benci V. Fortunato D.,}\textit{\ Three dimensional
vortices in Abelian Gauge Theories,} Nonlinear Analysis T.M.A. (2008).

\bibitem {befogranas}\textsc{V. Benci, D. Fortunato, } \emph{Solitary waves in
the nonlinear wave equation and in gauge theories}, J. Fixed Point Theory
Appl. \textbf{1} (2007), 61--86.

\bibitem {BF09TA}\textsc{Benci V. Fortunato D.,}\textit{Existence of
hylomorphic solitary waves in Klein-Gordon and in Klein-Gordon-Maxwell
equations, }Rendiconti dell'Accademia Nazionale dei Lincei, to appear (arXiv:0903.3508).

\bibitem {BFP98}\textsc{V. Benci, D. Fortunato, L. Pisani,} \textit{Soliton
like solution of a Lorentz invariant equation in dimension 3}, Reviews in
Mathematical Physics, \textbf{3} (1998), 315-344.

\bibitem {BGM09}\textsc{V. Benci, M. Ghimenti, A.M. Micheletti,} \emph{\ The
Nonlinear Schroedinger equation: solitons dynamics}, \textit{to appear }(arXiv:0812.4152).

\bibitem {bevi}\textsc{Benci V.}, \textsc{Visciglia N.}, \emph{Solitary waves
with non vanishing angular momentum}, Adv. Nonlinear Stud. \textbf{3} (2003), 151-160.

\bibitem {BL81}\textsc{H. Berestycki, P.L. Lions,} \textit{Nonlinear Scalar
Field Equations, I - Existence of a Ground State}, Arch. Rat. Mech. Anal.,
\textbf{82} (4) (1983), 313-345.

\bibitem {bus03}\textsc{Buslaev, Vladimir S.; Sulem, Catherine,} \textit{On
asymptotic stability of solitary waves for nonlinear Schr\"{o}dinger
equations. Annales de l'institut Henri Poincar\'{e} (C)} Analyse non
lin\'{e}aire, 20 no. 3 (2003), p. 419-475

\bibitem {ca}\textsc{Cassani D., }\textit{Existence and non-existence of
solitary waves for the critical Klein-Gordon equation coupled with Maxwell's
equations, }Nonlinear Anal.\textbf{58 }(2004), 733-747.

\bibitem {Ca03}\textsc{Cazenave,} \textsc{T.}\ \emph{Semilinear {S}%
chr\"{o}dinger equations}, Courant Lecture Notes in Mathematics, vol.~10, New
York University Courant Institute of Mathematical Sciences, New York, 2003.

\bibitem {CL82}\textsc{T.~Cazenave and P.L. Lions,} \emph{Orbital stability of
standing waves for some nonlinear {S}chr\"{o}dinger equations}, Comm. Math.
Phys. \textbf{85} (1982), no.~4, 549--561.

\bibitem {coleman78}\textsc{S. Coleman, V. Glaser, A. Martin,} \emph{Action
minima among solutions to a class of euclidean scalar field equation}, Commun.
Math. Phys. \textbf{58} (1978), 211--221

\bibitem {Coleman86}\textsc{S.Coleman,} \emph{\textquotedblleft
Q-Balls\textquotedblright}, Nucl. Phys. \textbf{B262} (1985), 263--283;
erratum: \textbf{B269} (1986), 744--745

\bibitem {cucc08}\textsc{Cuccagna, Scipio} \textit{On asymptotic stability in
3D of kinks for the }$\varphi^{4}$\textit{\ model.} Trans. Amer. Math. Soc.
360 (2008), no. 5, 2581--2614.

\bibitem {cucc08NS}\textsc{Cuccagna, Scipio; Mizumachi, Tetsu }\textit{On
asymptotically stability in energy space of ground states for nonlinear
Schr\"{o}dinger equations}. Comm. Math. Phys. 284 (2008), no. 1, 51--77.

\bibitem {tea}\textsc{D'Aprile T., Mugnai D.,}\textit{\ Solitary waves for
nonlinear Klein-Gordon-Maxwell and Schr\"{o}dinger -Maxwell equations, }Proc.
of Royal Soc. of Edinburgh, section A Mathematics, \textbf{134 }(2004), 893-906.

\bibitem {tea2}\textsc{D'Aprile T., Mugnai D.,}\textit{\ \ Non-existence
results for the coupled Klein-Gordon- Maxwell equations, }Advanced Nonlinear
studies, \textbf{4} (2004), 307-322.3.

\bibitem {D64}\textsc{C.H. Derrick,} \textit{Comments on Nonlinear Wave
Equations as Model for Elementary Particles}, Jour. Math. Phys. 5 (1964), 1252-1254.

\bibitem {GiNiNi}\textsc{B.Gidas, W.M.Ni, L.Nirenberg,} \textit{Symmetry and
related properties via the maximum principle}, Comm. Math. Phys., \textbf{68}
(1979), 209--243

\bibitem {gss87}\textsc{M. Grillakis, J. Shatah, W. Strauss,} \emph{Stability
theory of solitary waves in the presence of symmetry, I}, J. Funct. Anal.
\textbf{74} (1987), 160--197

\bibitem {Ka89}\textsc{Tosio Kato,} \emph{Nonlinear {S}chr\"{o}dinger
equations}, Schr\"{o}dinger operators (S\o nderborg, 1988), Lecture Notes in
Phys., vol. 345, Springer, Berlin, 1989, pp.~218--263.

\bibitem {kom96}Komech, A.; Vainberg, B. \textsc{On asymptotic stability of
stationary solutions to nonlinear wave and Klein-Gordon equations.} Arch.
Rational Mech. Anal. 134 (1996), no. 3, 227--248.

\bibitem {landaumeq}\textsc{Landau L.,Lifchitz E.}, \textit{M\'{e}canique},
Editions Mir, Moscow, 1966.

\bibitem {landau}\textsc{Landau L.,Lifchitz E.}, \textit{Th\'{e}orie du
Champ}, Editions Mir, Moscow, 1966.

\bibitem {ingl}\textsc{E. Long, }\textit{\textit{Existence and stability of
solitary waves in non-linear Klein-Gordon-Maxwell equations, }}Rev. Math.
Phys. \textbf{18} (2006), 747-779.

\bibitem {Poho}\textsc{Pohozaev S. I., }\textit{Eigenfunctions of the equation
}$\Delta u+\lambda f(u)=0,$Soviet Math. Dokl.,\textbf{165}, (1965) 1408-1412.

\bibitem {rosen68}\textsc{G. Rosen,} \emph{Particle-like solutions to
nonlinear complex scalar field theories with positive-definite energy
densities}, J. Math. Phys. \textbf{9} (1968), 996--998

\bibitem {rub}\textsc{Rubakov V., }\textit{Classical theory of Gauge fields,
}Princeton University press, Princeton 2002.

\bibitem {shatah}\textsc{J. Shatah,} \textit{Stable Standing waves of
Nonlinear Klein-Gordon Equations, }Comm. Math. Phys., 91, (1983), 313-327.

\bibitem {strauss}\textsc{W.A. Strauss}, \textit{Existence of solitary waves
in higher dimensions, }Comm. Math. Phys. \textbf{55 }(1977), 149-162

\bibitem {yangL}\textsc{\ Y. Yang,} \textit{Solitons in Field Theory and
Nonlinear Analysis, } Springer, New York, Berlin, 2000.
\end{thebibliography}
\end{document}